\documentclass[11pt,a4paper]{article}
 
\usepackage{amsmath}
\usepackage{amsfonts}
\usepackage{color}
\usepackage{cite}
\usepackage{graphicx}
\usepackage{float}
\usepackage{subfigure}

\numberwithin{equation}{section}

\title{\textbf{Wobbling kinks in a two-component scalar field theory: Interaction between shape modes}}

\author{A. Alonso-Izquierdo$^{(a,b)}$, D. Migu\'elez-Caballero$^{(c)}$, L. M. Nieto$^{(c)}$ and J. Queiroga-Nunes$^{(a,b)}$
\\ [1ex]
$^{(a)}$ Departamento de Matematica Aplicada, Universidad de Salamanca, \\ Casas del Parque 2, 37008 - Salamanca, Spain \\ [1ex]
$^{(b)}$  IUFFyM, Universidad de Salamanca, \\ Plaza de la Merced 1, 37008 - Salamanca, Spain \\ [1ex]
$^{(c)}$ Departamento de F\'{\i}sica Te\'orica, At\'omica y \'Optica, and IMUVA, \\
Universidad de Valladolid, 47011, Valladolid, Spain}

\date{\today}

\begin{document}

\maketitle

\begin{abstract}
In this paper the interaction between the shape modes of the wobbling kinks arising in the family of two-component MSTB scalar field theory models is studied. The spectrum of the second order small kink fluctuation in this model has two localized vibrational modes associated to longitudinal and orthogonal fluctuations with respect to the kink orbit. It has been found that the excitation of the orthogonal shape mode immediately triggers the longitudinal one. In the first component channel the kink emits radiation with twice the orthogonal wobbling frequency (not the longitudinal one as happens in the $\phi^4$-model). 
The radiation emitted in the second component has two dominant frequencies: one is three times the frequency of the orthogonal wobbling mode and the other is the sum of the frequencies of the longitudinal and orthogonal vibration modes. This feature is explained analytically using perturbation expansion theories.
\end{abstract}

\section{Introduction}

Topological defects have played an essential role over the last decades to describe phase transitions in cosmology \cite{Vilenkin1994, Vachaspati2006}, superconductivity in modern condensed matter \cite{Takashi2018}, protein folding \cite{Melnikov2019, Melnikov20192}, cluster of living species \cite{BazeiaCluster, BazeiaCluster2} and molecular systems \cite{Davydov1985, Bazeia1999}, just to quote a few application areas. Kinks are the simplest form of a topological defect solution in $(1+1)$-dimensions and they naturally arise in a variety of scalar field models, all of them sharing the essential feature of living in non-linear scenarios \cite{Manton2004, Shnir2018}. Two paradigmatic examples in this context are the sine-Gordon and the $\phi^4$-models. On the one hand, the sine-Gordon model is an integrable system that presents solitary wave solutions, called solitons. In fiber optics these solutions have been widely used to describe traveling digital signals over long haul fibers \cite{Mollenauer2006, Schneider2004, Agrawall1995}. On the other hand, the $\phi^4$-model is a non-integrable system that present solitary wave solutions called kinks. These solutions are stable under small fluctuations. In addition to the translational mode they have one localized vibrational mode, which is referred to as \textit{the shape mode}. In kink-antikink collisions energy can be transferred from one of these modes to the other. This \textit{resonant energy transfer mechanism} is responsible for the fractal structure found in the final versus initial velocity diagrams and the $n$-bounce window distribution, see \cite{Campbell1983, Sugiyama1979, Anninos1991, Malomed1992, Goodman2005, Goodman2008, Saadatmand2015, Takyi2016, Saadatmand2018, Adam2018, Gomes2018, Kevrekidis2019, Adam2019, Adam2019b, Adam2020, Mohammadi2020, Yan2020, Pereira2020, Manton2021, Blanco-Pillado2021, Adam2022}. This complex pattern arises in a wide variety of non-integrable models: the double sine-Gordon model \cite{Shiefman1979, Peyrard1983, Malomed1989,Gani1999, Goodman2002, Goodman2004, Gani2018, Simas2020},  the $\phi^6$-model \cite{Weigel2014,Gani2014,Marjaheh2017, Romanczukiewicz2017, Bazeia2018b, Lima2019},  the $\phi^8$-model  \cite{Christov2019b, Christov2019, Christov2020,Campos2021-1},  non-polynomial models \cite{Bazeia2017b, Bazeia2017a, Bazeia2019, Simas2016},  two-component scalar field theories \cite{Romanczukiewicz2008, Halavanau2012, Alonso2017, Alonso2018, Alonso2019, Ferreira2019, Alonso2020, Zhong2020, Alonso2021,Adam2021,Simas2022}, etc. In each of the previously cited works different aspects of the kink scattering are explored: resonance phenomena, long range forces, kink collisions with boundaries, impurities or defects, evolution of Bogomolnyi-Prasad-Sommerfield defects without intersoliton forces, presence of spectral walls, etc.

In the kink-antikink scattering, the wobbling kinks or wobblers (kinks whose vibrational or shape mode has been excited) play an essential role in the previous scattering processes because after the first collision the shape mode is excited and kinks becomes wobblers. For this reason, the evolution of single wobbling kinks has been thoroughly studied both analytically and numerically \cite{Getmanov1976, Segur1983, Manton1997, Barashenkov2009, Barashenkov2018}. In the $\phi^4$-model it has been found that while vibrating with wobbling frequency $\overline{\omega}=\sqrt{3}$ these solutions emit radiation with frequency $2\overline{\omega} = 2 \sqrt{3}$, that is, twice the natural vibrational frequency. This behavior has been analytically explained by employing perturbation theory in several works, see \cite{Segur1983,Manton1997,Barashenkov2009, Barashenkov2018}. For example, Manton and Merabet \cite{Manton1997} obtained the decay law of the wobbling amplitude by using a Lindstedt-Poincar\'e method. In the same context, Barashenkov and Oxtoby \cite{Barashenkov2009} employ a singular perturbation expansion which remains uniform to all orders by introducing a hierarchy of space and times scales. Furthermore, the collision between wobbling kinks has been investigated in the $\phi^4$-model \cite{Alonso2021b, Alonso2022} and the double sine-Gordon model \cite{Campos2021-2}.

A natural generalization of the $\phi^4$-model to theories involving several coupled scalar fields is given by the one-parameter family of the Montonen-Sarker-Trullinger-Bishop (MSTB) models. This system is a deformation of the $O(2)$ linear sigma model preserving the existence of two discrete vacua. It has been the focus of study by many researchers for decades \cite{Montonen1976, Rajaraman1975, Trullinger1976, Currie1979, Rajaraman1979, Subbaswamy1980, Subbaswamy1981, Magyari1984, Ito1985, Ito1985b, Alonso2000, Alonso2002c, Alonso2008}. A brief description of the main works concerning the family of MSTB models is introduced in \cite{Alonso2018b}. The usual kink (found in the one-component $\phi^4$-model) is embedded in these two-component scalar field theories. For the model parameter $\sigma$ in the interval $\sigma\in (1,\infty)$ the spectrum of the second-order small kink fluctuation operator implies the existence of a longitudinal vibrational mode (the usual shape mode found in the $\phi^4$-model) and an orthogonal vibrational mode, whose eigenfrequencies are, respectively, given by $\overline{\omega}=\sqrt{3}$ and $\widehat{\omega} = \sqrt{\sigma^2-1}$. This implies that the kink can vibrate in two different channels, periodically affecting the energy distribution of the extended particle described by this topological defect. In this work we will investigate the interaction between the vibrational modes of the kink in the MSTB model when the orthogonal shape mode is initially excited. It has been checked that in these circumstances the longitudinal shape mode is immediately triggered although it was initially unexcited. A surprising result is that the kink does not emit radiation with frequency $2\overline{\omega}$ in the longitudinal channel, as found in the $\phi^4$-model, but with frequency $2\widehat{\omega}$, that is, twice the frequency of the orthogonal shape mode. In the orthogonal channel, the frequencies of the emitted radiation are given by $\overline{\omega}+ \widehat{\omega}$ and $3 \widehat{\omega}$. In order to elucidate this behavior numerical and analytical approaches have been employed in this paper. This analysis can shed light on the energy transfer mechanism in scattering processes involving kinks in both one-component and two-component scalar field theory models.

The organization of this paper is as follows: in Section~\ref{MSTBmodel} the theoretical background of the MSTB model is introduced. The spectrum of the second order small kink fluctuation operator is thoroughly discussed. The eigenfunctions of the longitudinal and orthogonal shape modes are analytically identified. Section~\ref{Interaction} describes the problem addressed in this work: the evolution of a kink whose orthogonal shape mode has been initially excited. This problem is also numerically studied in this Section. The frequencies of the radiation reaching the simulation boundary are determined. The dependence of the amplitude of these eigenmodes on the model parameter $\sigma$ is also discussed.  
In the next two sections, the above numerical results are explained analytically by applying perturbation expansion theories.
In fact, in Section~\ref{MantonMerabet} the procedure  introduced by Manton and Merabet in \cite{Manton1997} is used, and in Section~\ref{BarashenkovOxtoby} the technique developed by Barashenkov and Oxtoby in \cite{Barashenkov2009} will be applied to our problem. Finally, the conclusions of this work are summarized in Section~\ref{conclusion}.

\section{The kink in the MSTB model: stability and two shape modes}\label{MSTBmodel}

The dynamics of the one-parameter family of MSTB models is governed by the action
\begin{equation}
S=\int d^2 x\,   \Big[ \frac{1}{2} \partial_\mu \phi \partial^\mu \phi+\frac{1}{2} \partial_\mu \psi \partial^\mu \psi - U(\phi,\psi) \Big] , \label{action}
\end{equation}
where the potential function $U(\phi,\psi)$ is determined by the fourth-degree polynomial
\begin{equation} 
U(\phi,\psi) = \frac{1}{2} (\phi^2 + \psi^2-1)^2 + \frac{1}{2} \sigma^2 \psi^2 \, . \label{potential}
\end{equation} 
As usual, $\phi,\psi:  \mathbb{R}^{1,1} \rightarrow \mathbb{R}$ are dimensionless real scalar fields and the Minkowski metric is taken as $(g_{\mu\nu})={\rm diag}\{1,-1\}$. Each member of this family of models is characterized by the value of the coupling constant $\sigma$ that appears in (\ref{potential}), which belongs to the interval $\sigma \in [0,\infty)$. The field equations in this case are given by  the following system of coupled nonlinear partial differential equations  of the nonlinear Klein-Gordon type:
\begin{eqnarray}
\frac{\partial^2 \phi}{\partial t^2} - \frac{\partial^2 \phi}{\partial x^2} \!\!&\!\!=\!\!&\!\! 2 \phi(1-\phi^2 - \psi^2) \, , \label{pde1} \\
\frac{\partial^2 \psi}{\partial t^2} - \frac{\partial^2 \psi}{\partial x^2} \!\!&\!\!=\!\!&\!\! 2 \psi \Big(1-\phi^2 - \psi^2 - \frac{1}{2} \sigma^2\Big) \, . \label{pde2} 
\end{eqnarray}
Solutions of the equations (\ref{pde1}) and (\ref{pde2}) involves a kink variety with a very rich structure. The usual kink solution that arises in the $\phi^4$-model
\begin{equation}
\phi_K(x) = \tanh {x} \label{kinkdef}
\end{equation}
and its antikink are embedded into this two-component scalar field theory model as
\begin{equation}
K^{(\pm)}(x) = (\pm \, \phi_K(x) \,\, , \,\, 0)^t = (\pm \, \tanh {x} \,\, , \,\, 0)^t  .\label{kink}
\end{equation}
Here and throughout this article we assume that the center of the kink is at the origin, $x=0$, but it could be located at another point $x_C\in \mathbb{R}$, which would be achieved with a simple translation in \eqref{kinkdef}.
The signs in (\ref{kink}) distinguish between the kink ($+$) and the antikink ($-$). The solution (\ref{kink}) arises for any value of the model parameter $\sigma$ although this topological defect is unstable for $\sigma<1$, as we will see below. In this regime, a pair of two nonzero component topological kinks arise, which are less energetic than the kinks (\ref{kink}) and become linearly stable. In addition to these solutions, a one-parametric family of non-topological kinks emerges for each vacua in the model, see \cite{Ito1985, Ito1985b, Alonso2018b}. In this paper, however, we are interested in the regime $\sigma\geq 1$, where the one nonzero component kink (\ref{kink}) is the only topological defect and is stable. This statement can be checked by analyzing the evolution of a perturbation of the static kink (\ref{kink}) of the form
\begin{equation}
\widetilde{K}(x,t;\omega,a) = K^{(\pm)}(x) + a \, e^{i\omega t} \, F_\omega (x) ,\label{perturbation}
\end{equation}
where $a$ is a small real parameter. This leads to the spectral problem
\begin{equation}
{\cal H} \, F_\omega(x) = \omega^2 F_\omega(x), \label{spectralproblem}
\end{equation}
where ${\cal H}$ is the second order small fluctuation operator
\begin{equation}
{\cal H} = \left( \begin{array}{cc} {\cal H}_{11} & 0 \\0 & {\cal H}_{22} \end{array} \right) =\left( \begin{array}{cc}
-\frac{d^2}{dx^2} + 4 -6 \, {\rm sech}^2 {x} & 0 \\ 0 & -\frac{d^2}{dx^2} + \sigma^2 -2 \, {\rm sech}^2 {x}
\end{array} \right)  \, . \label{operator}
\end{equation} 
From the expression (\ref{perturbation}) it is clear that the solution (\ref{kink}) is stable only if the eigenvalues of the operator (\ref{operator}) are not negative. Note that (\ref{operator}) is a diagonal matrix differential operator. The spectral problem (\ref{spectralproblem}) in this case consists of two exactly solvable spectral problems (independent of each other) corresponding to Sch\"odinger operators ${\cal H}_{11}$ and ${\cal H}_{22}$ with P\"oschl-Teller potential wells. Taking into account that the orbit of the kink solution (\ref{kink}) is located on the $\phi$-axis of the internal plane, perturbations of the form 
\[
\overline{F}_\omega(x) = (\overline{\eta}(x),0)^t
\]
characterize longitudinal fluctuations on the kink while those of the form 
\[
\widehat{F}_\omega (x)=(0,\widehat{\eta}(x))^t
\]
determine orthogonal or transverse fluctuations of (\ref{kink}). As mentioned above, these types of perturbations are decoupled in this case, so that the spectrum of (\ref{operator}) is described as follows:
\begin{itemize}
\item \textit{Longitudinal eigenmodes:} It can be checked that the operator (\ref{operator}) involves two discrete longitudinal eigenfunctions. These states are the zero mode
\begin{equation}
\overline{F}_0(x)= (\overline{\eta}_0(x),0)^t = ({\rm sech}^2 x , 0)^t  \label{zeromode1}
\end{equation}
and the so-called shape mode
\begin{equation}
\overline{F}_{\sqrt{3}} \,(x)= (\overline{\eta}_D(x),0)^t= ({\rm sech}\, x \, \tanh x , 0)^t \label{shapemode1}
\end{equation}
whose eigenvalue is given by $\overline{\omega}^2 = 3$. This means that the kink when excited by this longitudinal shape mode vibrates with frequency $\overline{\omega}=\sqrt{3}$. It can be checked that the maximum deviation of this vibrating kink (or wobbler) from the static kink (\ref{kink}) happens at the points
\begin{equation}
x_M^{(\pm)} = \pm \, {\rm arccosh}\,\sqrt{2} \, . \label{xm}
\end{equation}
The longitudinal continuous spectrum emerges on the threshold value $\overline{\omega}_c^2 = 4$, that is, $\overline{\omega}_{q}^2 = 4 + q^2$ with $q\in \mathbb{R}$. The corresponding eigenfunctions read
\begin{equation}
\overline{F}_{\sqrt{4+q^2}} (x) = (\overline{\eta}_{q}(x),0)^t = \Big( e^{iqx} [-1-q^2+ 3 \, \tanh^2 x -3 i q \tanh x]\, , 0 \,\, \Big)^t \, . \label{continuous1}
\end{equation}
The relations (\ref{zeromode1}), (\ref{shapemode1}) and (\ref{continuous1}) are also employed to define respectively the functions $\overline{\eta}_0(x)$, $\overline{\eta}_D(x)$ and $\overline{\eta}(x)$. Obviously, the spectrum of the longitudinal fluctuations is the same as that found for the one-component kink in the $\phi^4$-model because, as previously mentioned, this kink is embedded in the MSTB model. Indeed, $\overline{\eta}_D(x)$ is the usual shape mode that turns a static kink into a wobbling kink, see \cite{Segur1983, Manton1997,Barashenkov2009, Barashenkov2018 ,Alonso2021b, Alonso2022}. As we will see below, in the MSTB model the kink can also vibrate in the orthogonal direction through a vibrational mode, which we will continue to call \textit{shape mode}.

\item \textit{Orthogonal eigenmodes:} In this case, the spectrum of orthogonal fluctuations contains only a discrete eigenvalue $\widehat{\omega}^2 = \sigma^2 -1$, whose eigenfunction is
\begin{equation}
\widehat{F}_{\sqrt{\sigma^2-1}} (x) = (0,\widehat{\eta}_D(x))^t = (0 , {\rm sech}\,x)^t\, .  \label{shapemode2}
\end{equation}
This orthogonal vibrational or shape mode maximally vibrates at the point 
\begin{equation}
x_0=0 \label{x0}
\end{equation}
with frequency $\widehat{\omega} =\sqrt{\sigma^2-1}$ if $\sigma>1$. Note that if $\sigma<1$ the kink (\ref{kink}) becomes unstable because the eigenvalue $\widehat{\omega}^2$ is negative. In this case the kink (\ref{kink}) decays to one of the previously mentioned less energetic two non-null component kinks which arise in this regime. An interesting remark is that for $\sigma=1$ the solution (\ref{kink}) has two zero modes despite the fact that it is the only kink present in the model and, therefore, it is not a member of a one-parametric family of solutions. This is a very singular situation. Of course, this model parameter value defines a phase transition where a family of non-topological kinks starts to emerge, see \cite{Alonso2018b}. 

Finally, the orthogonal continuous spectrum verifies the dispersion relation 
$\widehat{\omega}_{q}^2 = \sigma^2 + q^2$, $q\in \mathbb{R}$, with eigenfunctions
\begin{equation}
\widehat{F}_{\sqrt{\sigma^2+ q^2}} (x) = (0,\widehat{\eta}_{q}(x) )^t = \left( 0 \,\,,\,\, e^{i qx} (q+i \, \tanh x)\right)^t \, , \label{continuous2}
\end{equation}
which emerge on the threshold value $\widehat{\omega}_c^2 = \sigma^2$. 
\end{itemize}

\begin{figure}[htb]
\centering
\subfigure[Longitudinal fluctuations.]{\label{fig:espectro1a}
\includegraphics[width=0.452\textwidth]{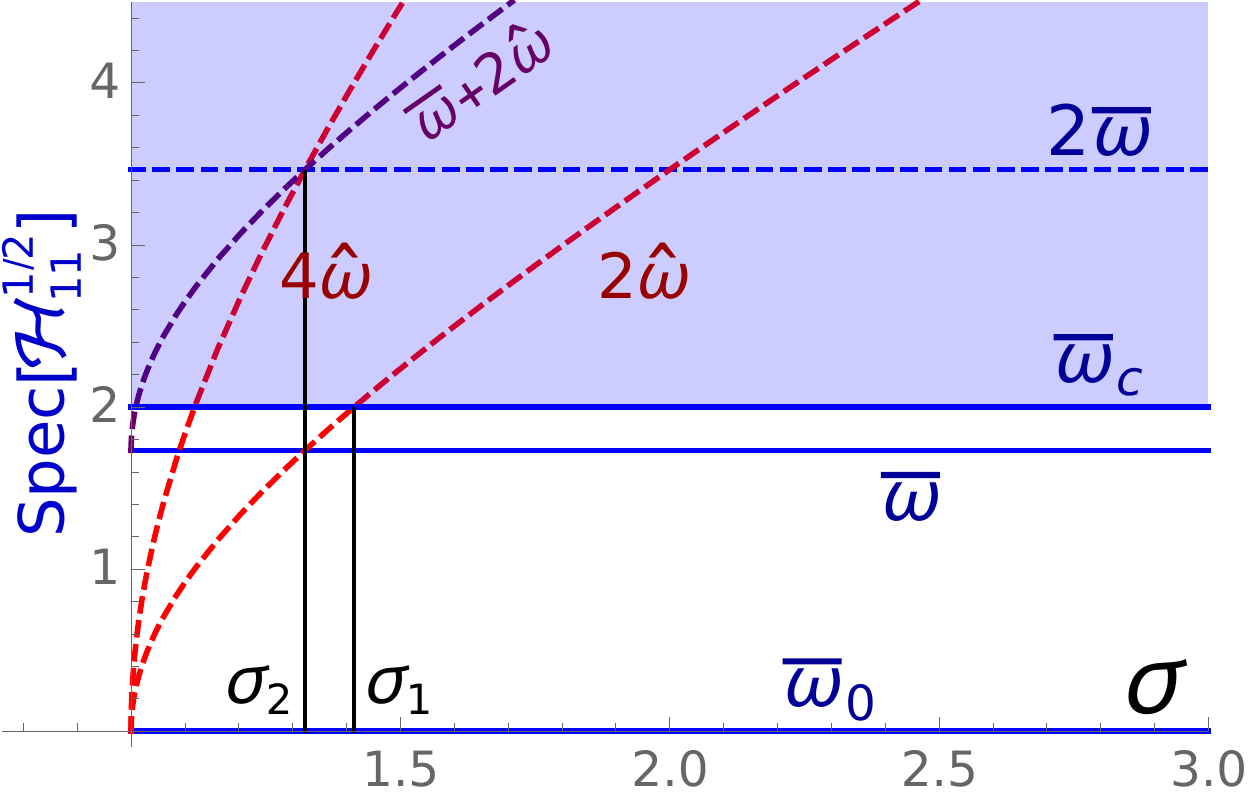}}
\qquad\quad
\subfigure[Orthogonal fluctuations.]{\label{fig:espectro1b}
\includegraphics[width=0.452\textwidth]{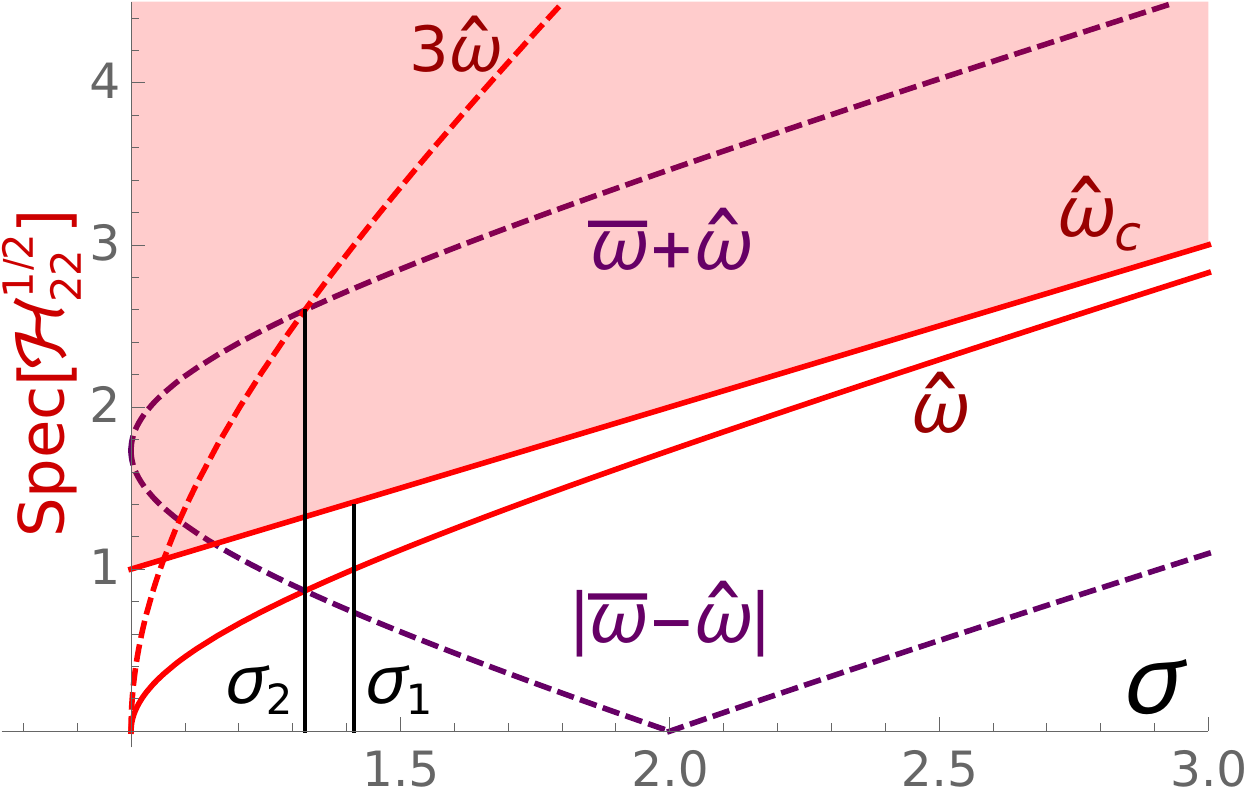} }
	\caption{\small Solid lines  are the eigenfrequencies $\omega$ (the square root of the eigenvalues of the operator (\ref{operator})) associated to the longitudinal (a) and orthogonal (b) fluctuations as a function of the coupling constant $\sigma$. Some combinations of these eigenfrequencies that will be relevant later are also included (dashed lines).} 
	\label{fig:espectro1}
\end{figure}

\noindent The structure of the spectrum of \eqref{spectralproblem}--\eqref{operator} depends on the value of the coupling constant $\sigma$. In Figure~\ref{fig:espectro1} the eigenfrequencies $\omega$ derived from the spectrum of the operator (\ref{operator}) are represented as functions of the model parameter $\sigma$ by solid lines. The longitudinal eigenfrequencies are shown in Figure \ref{fig:espectro1a} while the orthogonal ones are shown in Figure \ref{fig:espectro1b}. For later purposes, some discrete eigenfrequencies combinations have also been represented in Figure \ref{fig:espectro1} by dashed curves. In particular, the values $2\overline{\omega}$, $2\widehat{\omega}$, $4\widehat{\omega}$ and $\overline{\omega}+2\widehat{\omega}$ are included in the first graph and $\overline{\omega}+\widehat{\omega}$, $|\overline{\omega}-\widehat{\omega}|$ and $3\widehat{\omega}$ in the second. This allows us to distinguish two significant values of $\sigma$, which will play a relevant role later. Specifically we have the following:
\begin{itemize}

\item 
$\sigma_1=\sqrt{2} \approx 1.41421$. For $\sigma>\sigma_1$ the relation $2\widehat{\omega} > \overline{\omega}_c$ holds, that is, twice the frequency $\widehat{\omega}$ of the orthogonal shape mode can be found as a eigenfrequency in the continuous longitudinal  spectrum, see Figure~\ref{fig:espectro1a}. On the other hand, if $\sigma<\sigma_1$ the inequality $2\widehat{\omega} < \overline{\omega}_c$ holds, which means that there is no radiation with this frequency in the first component. 

\item 
$\sigma_2=\frac{1}{2} \sqrt{7} \approx 1.32288$. For $\sigma = \sigma_2$ the equality $2\widehat{\omega} = \overline{\omega}$ is verified, that is, twice the frequency of the orthogonal shape mode coincides with the frequency of the longitudinal shape mode. Also note that $3\widehat{\omega}  = \overline{\omega} + \widehat{\omega}$ and $\widehat{\omega}  = \overline{\omega} - \widehat{\omega}$ are also valid, see Figure~\ref{fig:espectro1b}. 

\end{itemize}

It is worth anticipating that these values of $\sigma$ will become especially important later on because just in $\sigma_1$ and $\sigma_2$ there are divergences in the radiation amplitudes that will be calculated in Sections~\ref{MantonMerabet} and~\ref{BarashenkovOxtoby}.

\section{Interaction between the normal modes of vibration}\label{Interaction}

In this section we will analyze the interaction between the shape modes described above, which allow the kink solution (\ref{kink}) to vibrate in a longitudinal or orthogonal channel. This interaction is governed by the nonlinear Klein-Gordon equations (\ref{pde1})--(\ref{pde2}). 
The coupling between the field components present in the different terms of the MSTB potential (\ref{potential}) implies that the excitation of one of the eigenmodes can excite other different vibrational or radiation modes. 
For our case, it is clear that orthogonal modes cannot be triggered by longitudinal modes. If the second field component vanishes, $\psi=0$, then the problem reduces to the usual one-component $\phi^4$-model and the second component $\psi$ is not affected by kink evolution. 
Here, we will address the reverse situation, that is, we will discuss how the longitudinal modes are affected when the orthogonal shape mode $\widehat{F}(x)$ of the static kink (\ref{kink}) is excited. In this scenario, the initial configuration can be characterized by the linear approximation 
\begin{equation}
\widetilde{K}(x,t; a) = K^{(+)}(x) + a \, \sin (\widehat{\omega} t) \, 
\widehat{F}_{\sqrt{\sigma^2-1}}(x) 
\label{initial01}
\end{equation}
derived from linear stability analysis. By construction, this is a good approximation when $a$ is small. Using \eqref{kink} and \eqref{shapemode2}, the expression (\ref{initial01}) can be written in components as 
\begin{equation}
\widetilde{K}(x,t; a) =\Big(\phi_K (x) \, , \, a \, \sin (\widehat{\omega} t) \,\, \widehat{\eta}_{D} ({x}) \Big)^t
= \Big(\tanh {x} \, , \, a \, \sin (\sqrt{\sigma^2-1}\, t ) \,\, {\rm sech}\, {x} \Big)^t \, .
\label{initial02}
\end{equation}
After the excited kink (\ref{initial01})--(\ref{initial02}) has evolved sufficiently, we expect other eigenmodes to become excited, so that eventually the kink is more accurately described by the expression\footnote{Recall that for the usual $\phi^4$-model, the kink evolves by vibrating through the initially excited shape mode channel while emitting radiation with twice the $\overline{\omega}$ wobbling frequency. Using the notation employed in (\ref{final01}) this would mean  $a_{\overline{\omega}}\neq 0$ and $a_{2\overline{\omega}}\neq 0$. Note that in this case the frequency $2\overline{\omega}$ is embedded in the continuous spectrum.}:
\begin{equation}
\widetilde{K}(x,t; a) = K^{(+)}(x) + \sum_{\omega \in {\rm Spec}\,{\cal H}} a_\omega \, \sin (\omega t+\varphi_\omega) \, F_\omega (x) , \quad \mbox{for} \quad t\gg 0
\label{final01}
\end{equation}
which is a generalization of \eqref{perturbation} and \eqref{initial01}.

\begin{figure}[htb]
	\centerline{\includegraphics[width=0.41\textwidth]{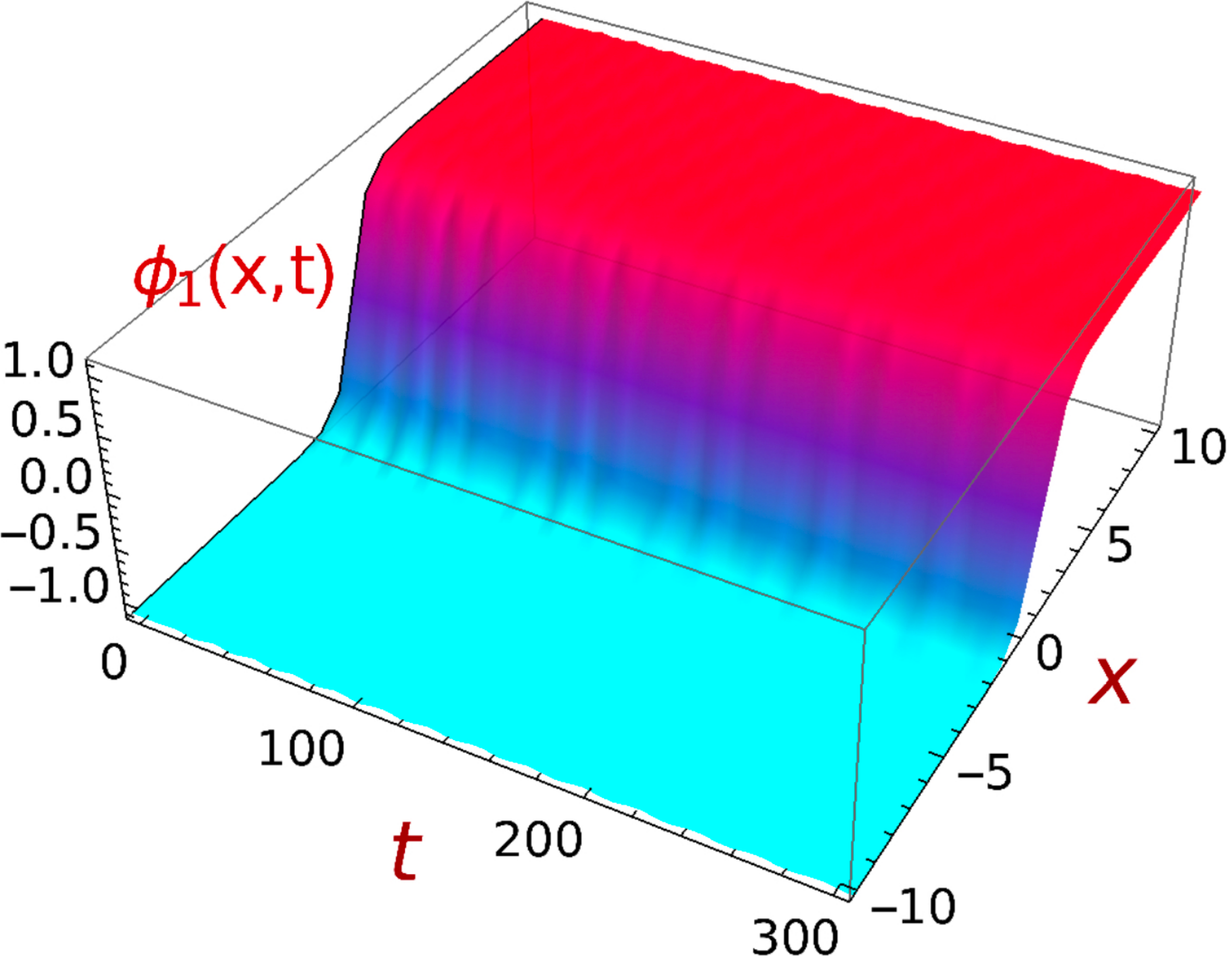} 
	\qquad\qquad  
	\includegraphics[width=0.4\textwidth]{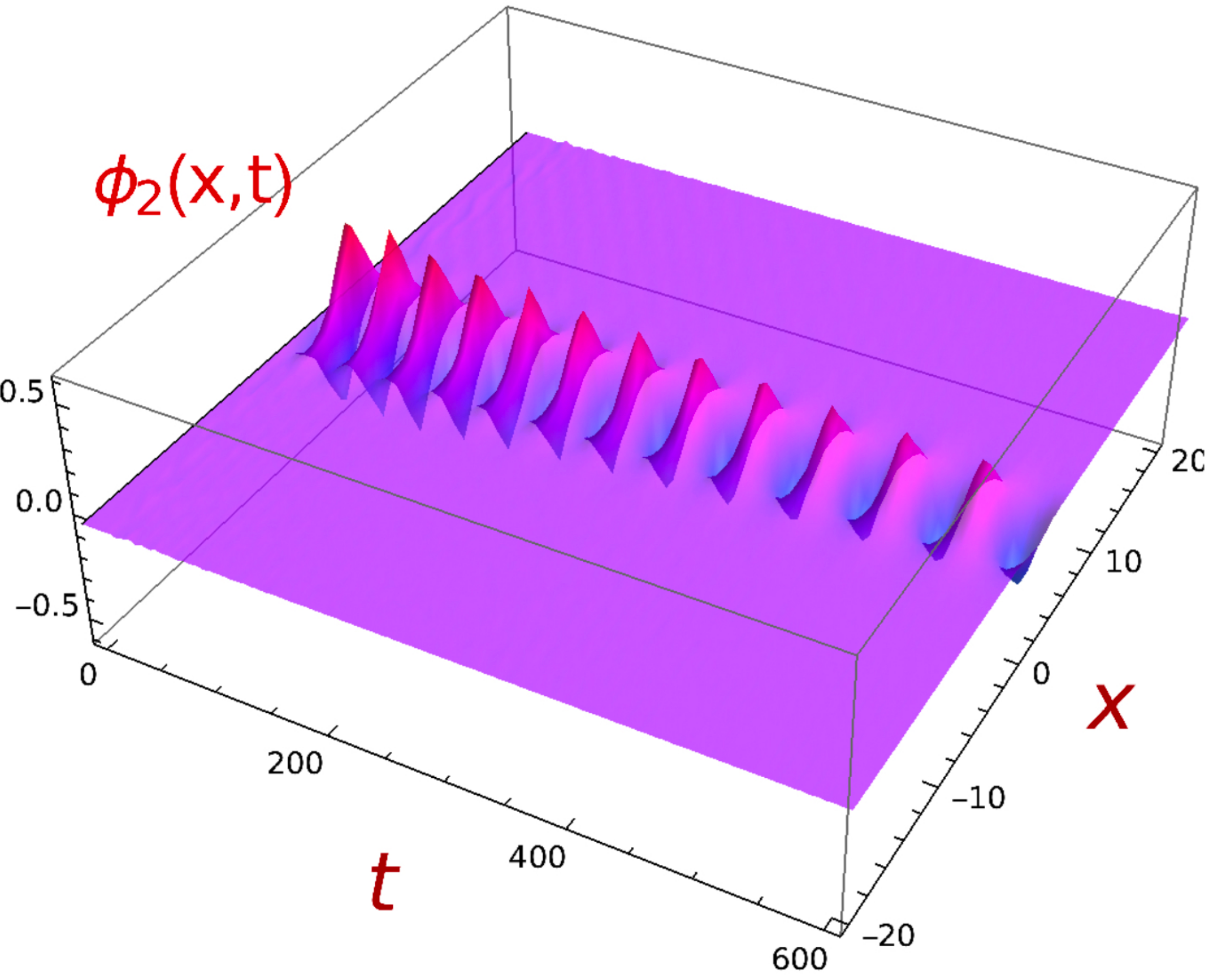} }
	\caption{\small Evolution of the first (left) and second (right) field components of the kink (\ref{kink}) initially excited by the orthogonal shape mode $\widehat{F}(x)$ (\ref{shapemode2}) with amplitude $a=0.6$.} \label{fig:simulacion}
\end{figure}

In the following sections we will study the problem proposed in this work from two different perspectives. On the one hand, we will use numerical analysis to extract the spectral information associated with the excitation of the different eigenmodes of the evolving kink. 
For example, Figure~\ref{fig:simulacion} shows the evolution of the two-component kink solution when the orthogonal shape mode has been initially excited with amplitude $a=0.6$. 
The value of $a$ has been chosen high enough to better visualize the behavior of the fields in the graphics. 
We can  clearly see that the initial excitation of the orthogonal shape mode $\widehat{F}(x)$ causes an almost immediate excitation of the longitudinal shape mode $\overline{F}(x)$. 
Note the small ripples around the center of the kink in the first plot. Another characteristic that is shown in Figure~\ref{fig:simulacion} is that the amplitude of the orthogonal shape mode decreases as time passes. 
As  mentioned above, this behavior is expected since this  also occurs in the evolution of the wobbling kink in the one-component $\phi^4$-model. As a consequence, some radiation is emitted in this process. 
As stated above, in the $\phi^4$ model plane waves are generated with a frequency of $2 \overline{\omega}$. In our model, radiation can propagate through both components of the field.
Furthermore, the kink (\ref{kink}) can vibrate with two different wobbling frequencies $\overline{\omega}$ and $\widehat{\omega}$. 
For this reason it is difficult to anticipate the frequencies of the radiation emitted by the excited kink (\ref{initial01}). 
On the other hand, in Section~\ref{MantonMerabet} an analytical approach based on the perturbation expansion theory will be introduced.
We will use two different techniques that have been employed by Manton and Merabet \cite{Manton1997} and by Barashenkov and Oxtoby \cite{Barashenkov2009} in the analysis of the wobbling kink in the $\phi^4$-model. 
This study will be applied to the predominant regime $\sigma \in (\sigma_1,\infty)$, see Figure~\ref{fig:espectro1}. This will allow us to obtain an analytical understanding of the numerical results and of the mechanism involved in the interaction between the different eigenmodes. 
The  $\sigma \in (1,\sigma_1)$ regime implies much more complex dynamics, where added frequencies start to come into play due to couplings between higher order form modes.

\subsection{Numerical simulations}

As  mentioned before, in this section we will perform numerical simulations to gain a global understanding of the problem. This is characterized by a static kink whose orthogonal shape mode is initially excited with a certain initial amplitude $a$ and evolves fulfilling the partial differential equations (\ref{pde1}) and (\ref{pde2}). 
These equations will be discretized  using an explicit fourth-order  finite difference algorithm implemented with fourth-order Mur boundary conditions, which has been designed to deal with non-linear Klein-Gordon equations (see the Appendix in \cite{Alonso2021}). The initial conditions are easily implemented from the initial configuration (\ref{initial02}) with $t=0$. 
The simulations have been executed in the spatial interval $x\in [-100,100]$ for a time range $t\in [0,1200]$. The space and time intervals have been chosen to correctly estimate the spectral data associated with the vibrations of the fields using a fast Fourier transform algorithm. 
Different choices of  space and time intervals have also been considered giving similar results. The  spectral analysis mentioned above has been applied to time series obtained by evaluating the field components of the solutions at different  points in space. This procedure is justified by the fact that the shape modes are located and the estimation of their frequencies and amplitudes is optimized using different points $x_i$. The choice of these points is specified as follows:

\begin{enumerate}
	\item Spectral information for time series evaluated at $x=x_0=0$. The vibration of the orthogonal shape mode $\widehat{F}(x)$ is most pronounced at the origin of the spatial axis, see (\ref{shapemode2}). The time series extracted by evaluating the second field at this point allows us to assess, for example, the resulting amplitude of the orthogonal shape mode $\widehat{F}(x)$ associated with the evolving kink. Note that the first field component in this solution always vanishes at $x_0$. 

	\item Spectral information for time series evaluated at $x=x_M^{(+)}$. The amplitude of the longitudinal shape eigenmode is maximized at these points, see (\ref{shapemode1}) and (\ref{xm}). The spectral analysis at the point $x_M^{(+)}$ defines an optimized method to asses the excitation of this vibration channel. In contrast to the point $x_0$, which can be considered the source of the radiation emitted by the vibrating kink, some continuous eigenmodes propagating from the origin can be detected here.   
	
	\item Spectral information for time series evaluated at $x=x_B$ with $x_B$ representing a spacial value near the limits of the simulation. The set of radiation modes emitted by the wobbling kink can be identified by the fluctuations away from the kink center $x_C$. This makes it possible to identify the frequencies of the radiation emitted in both the longitudinal and orthogonal channels.
\end{enumerate}

The spectral analysis of the time series corresponding to the points $x_0$, $x_M^{(+)}$ and $x_B$ is shown in the following figures for various values of the coupling constant $\sigma$ and for the value of the initial amplitude $a=0.1$. 
It has been verified that for other values of $a$ in the range $(0,0.6]$ the results are very similar. For this reason the particular value $a=0.1$ can be considered as a representative case of the problem we are dealing with. 
Obviously, the resulting spectral information depends on the value of the model parameter $\sigma$. In the following figures, the results for the particular values $\sigma=1.2$, $\sigma=1.5$ and $\sigma=2.5$ are illustrated. These cases describe the wide variety of behaviors that arise in this problem. Some of the most interesting features are set out in the following list:

\begin{itemize}
\item 
 Figure \ref{fig:espectro2} represents the amplitude $a_\omega$ of the different longitudinal eigenmodes $\overline{F}_\omega(x)$ as a function of the frequency $\omega$ when the time series are evaluated at the point $x_M$, which is the optimum point to assess the excitation of the longitudinal shape mode $\overline{F}_{\overline{\omega}}(x)$. 
	You can see that this mode is excited even though it initially was not. The first field component evaluated at this point also involves a vibration with frequency $2\widehat{\omega}$. To check if it is a localized fluctuation or corresponds to a radiation mode, the spectral analysis must be carried out far from the  center of the kink. This is discussed in the next point.

\begin{figure}[htb]
	\centerline{\includegraphics[width=0.3\textwidth]{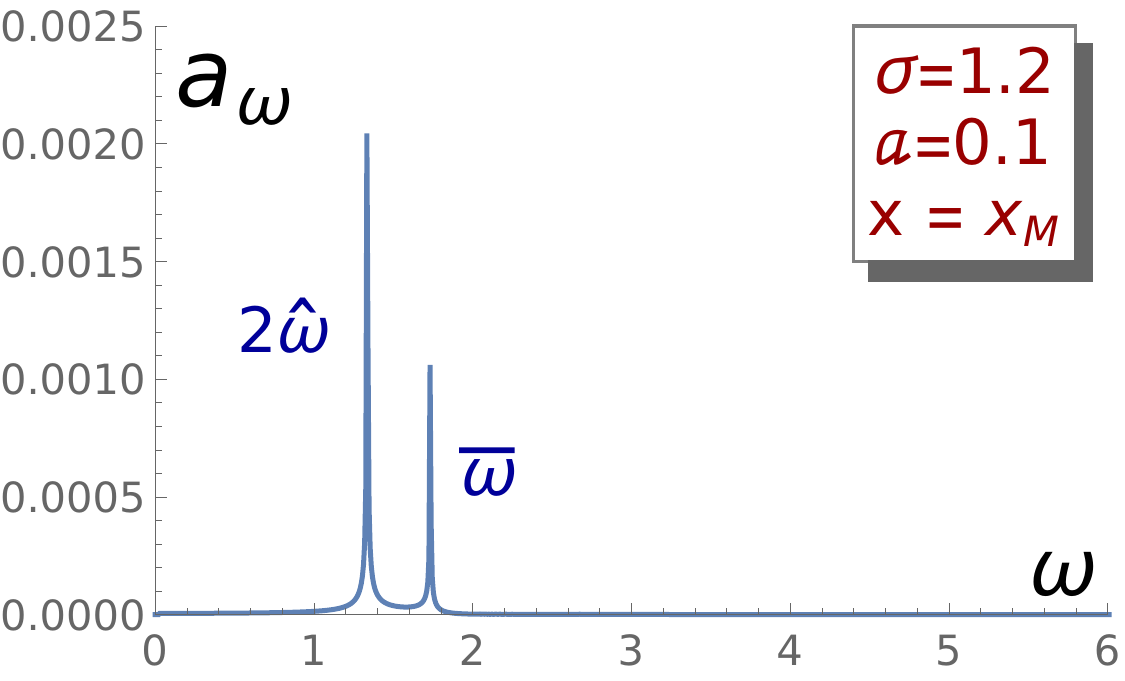} 
	\qquad 
	\includegraphics[width=0.3\textwidth]{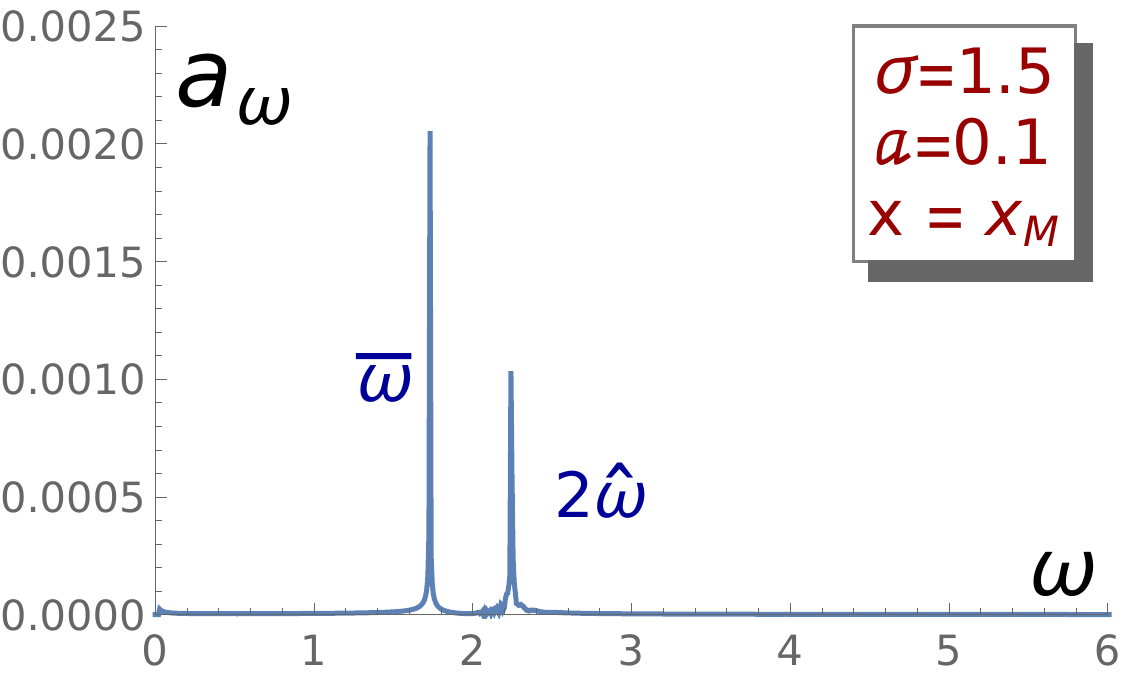} 
	\qquad
	\includegraphics[width=0.3\textwidth]{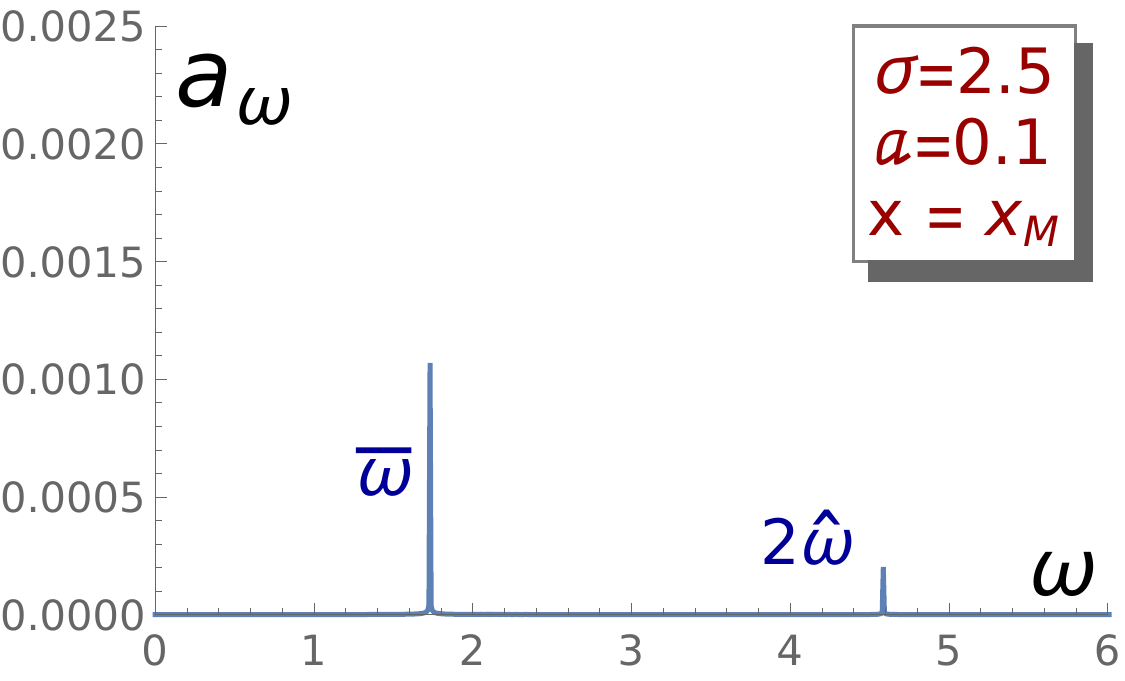}}
	\caption{\small Spectral distribution of the time series obtained by evaluating the first field component at the point $x_M^{(+)}$ for the evolving kink with initial orthogonal shape mode amplitude $a=0.1$. This information is displayed for three representative values of the model parameter: $\sigma=1.2$ (left), $\sigma=1.5$ (middle), and $\sigma=2.5$ (right).} 
	\label{fig:espectro2}
\end{figure}

\item 
Figure \ref{fig:espectro3} shows the spectral distribution of the radiation emitted in the first component channel that reaches the limits of the simulations. In practice, the time series in this case are constructed at a point $x_B$ (far from the  center of the kink). 
In the one-component $\phi^4$-model, nonlinearity causes the emission of radiation with frequency $2 \overline{\omega}$.  
An unexpected novelty appears in our model: radiation is also emitted but its frequency is not $2 \overline{\omega}$ but $2\widehat{\omega}$, that is, twice the frequency of the orthogonal shape mode. 
This behavior can be observed in Figure \ref{fig:espectro3} (center and right) for the model parameters $\sigma=1.5$ and $\sigma=2.5$.  
Note that the frequency $2\widehat{\omega}$ is embedded in the longitudinal continuous spectrum for $\sigma > \sigma_1= \sqrt{2}$ which includes the cases mentioned above. 
\begin{figure}[htb]
	\centerline{\includegraphics[width=0.3\textwidth]{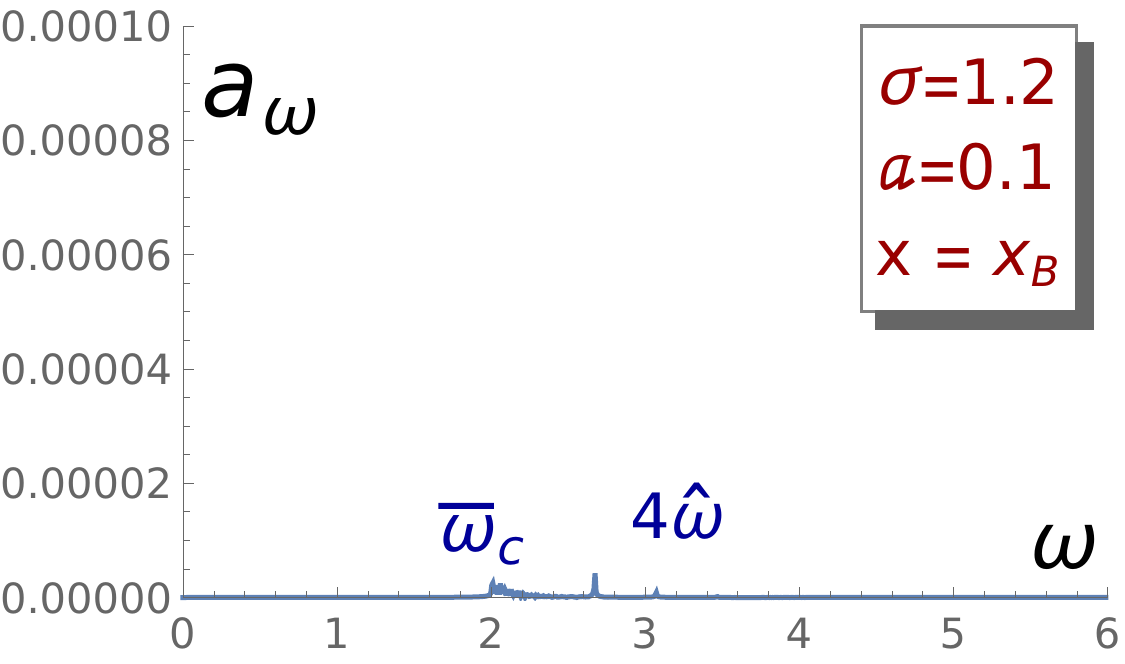} 
	\qquad
	\includegraphics[width=0.3\textwidth]{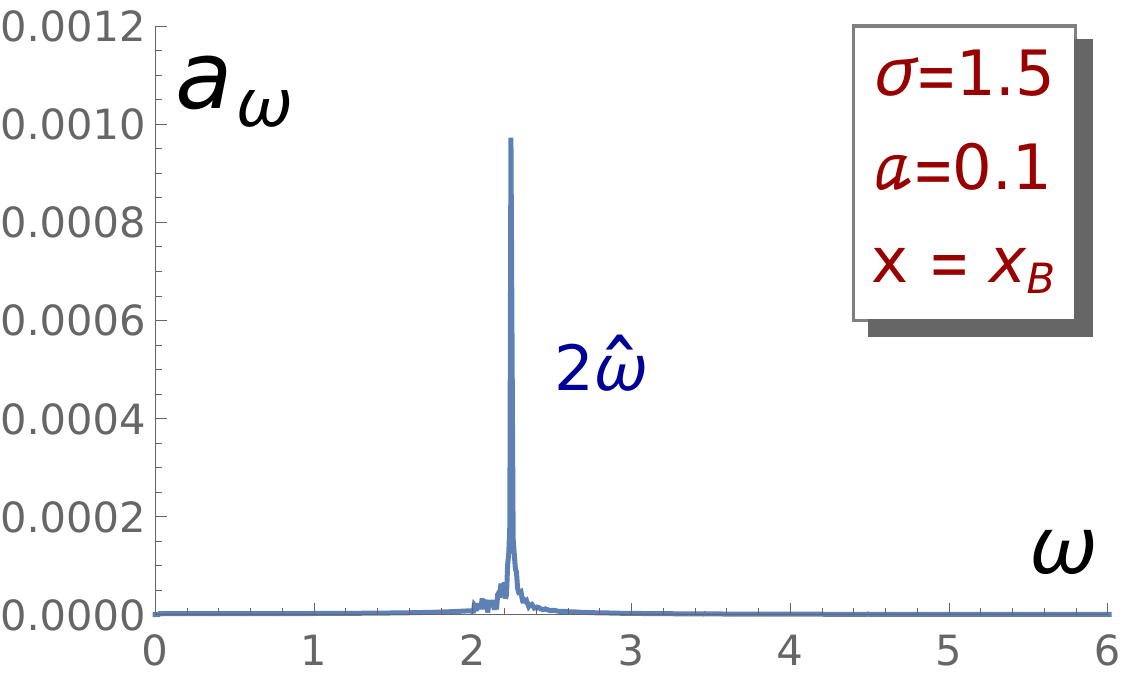} 
	\qquad
	\includegraphics[width=0.3\textwidth]{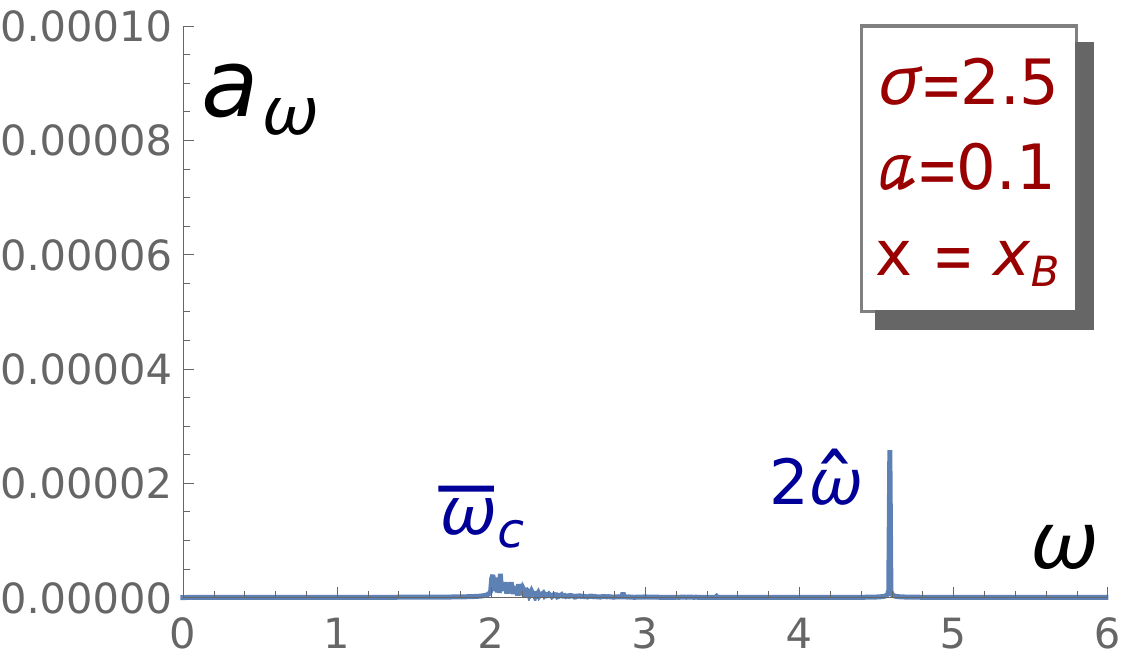}}
	\caption{\small Spectral distribution of the time series obtained by evaluating the first field component at the point $x_B$ for the evolving kink with initial orthogonal shape mode amplitude $a=0.1$. This information is shown for three representative values of model parameter: $\sigma=1.2$ (left), $\sigma=1.5$ (middle) and $\sigma=2.5$ (right).} 
	\label{fig:espectro3}
\end{figure}
The amplitude of the excited eigenmodes depends on the value of the model parameter $\sigma$. This can be partly explained by the fact that higher frequencies are more difficult to excite and that the eigenvalues associated with the orthogonal eigenfluctuations depend on $\sigma$. 
The case $\sigma=1.2$ shown in Figure \ref{fig:espectro3} (left) does not satisfy the previous inequality, so there is no continuous longitudinal  eigenfunction with frequency $2\widehat{\omega}$. Instead, a continuous eigenmode with frequency $4\widehat{\omega}$  can be identified, albeit very slightly excited. 
Clearly, this indicates that higher order nonlinear terms are involved in the emission of this radiation. 
On the other hand, note that the frequency $\overline{\omega}$ observed in Figure \ref{fig:espectro2} is not present here because this massive eigenmode is localized around the kink center.

\item The pattern found in Figure \ref{fig:espectro4} is different. The spectral analysis is applied, in this case, on time series obtained by the values of the second field component extracted from our simulations computed at the kink center $x_0=0$. By symmetry, this point can be considered the source of the radiation emitted by the wobbling kink. For this reason, the plots shown in Figure \ref{fig:espectro4} show that the orthogonal shape mode (initially excited by construction) continues to be excited over time. However, the amplitude of this localized vibration seems to follow a non-trivial behavior. We will return to this point later.

\begin{figure}[htb]
	\centerline{\includegraphics[width=0.3\textwidth]{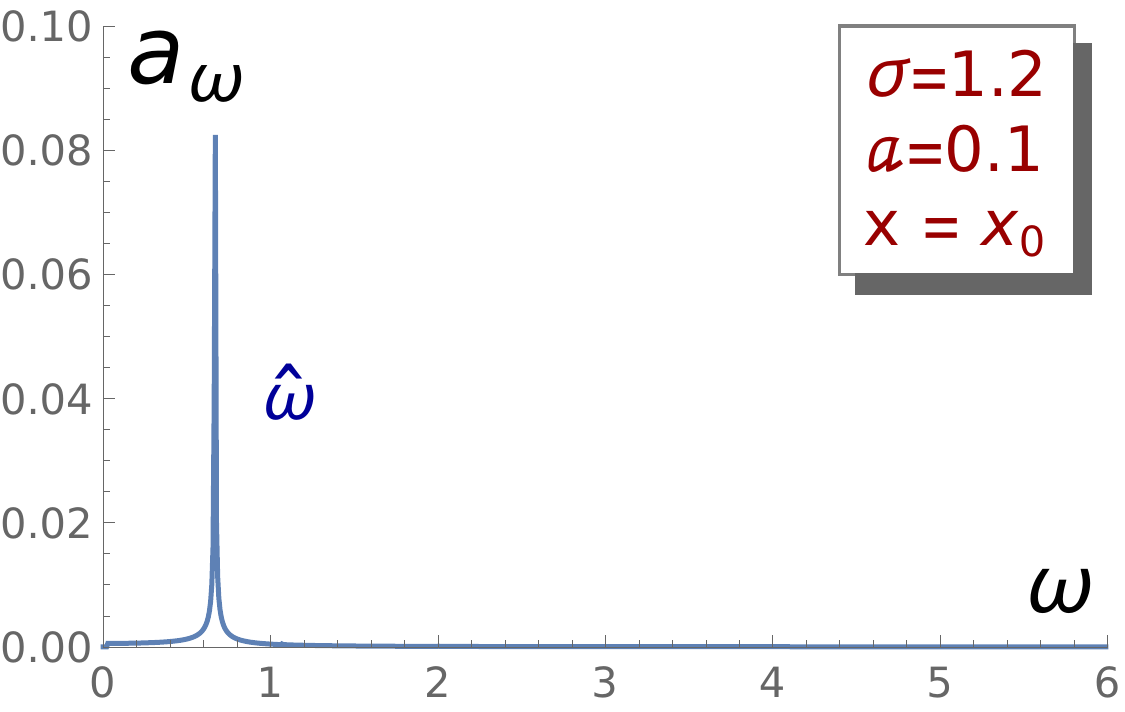} 
	\qquad	
	\includegraphics[width=0.3\textwidth]{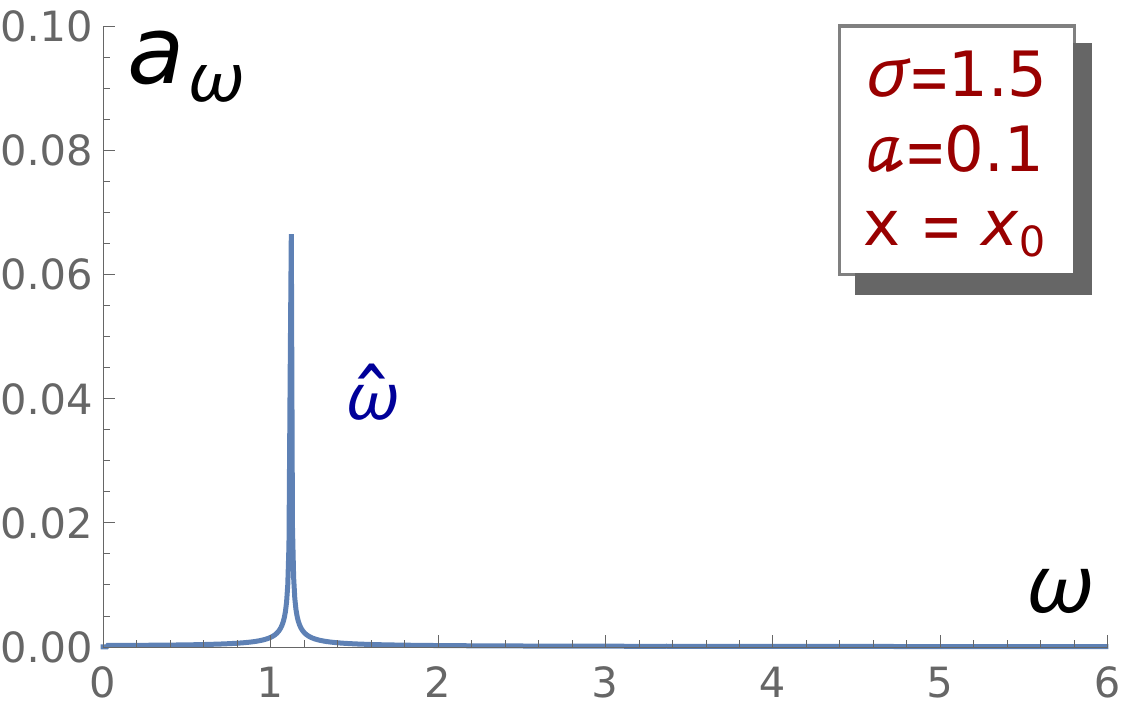} 
	\qquad
	\includegraphics[width=0.3\textwidth]{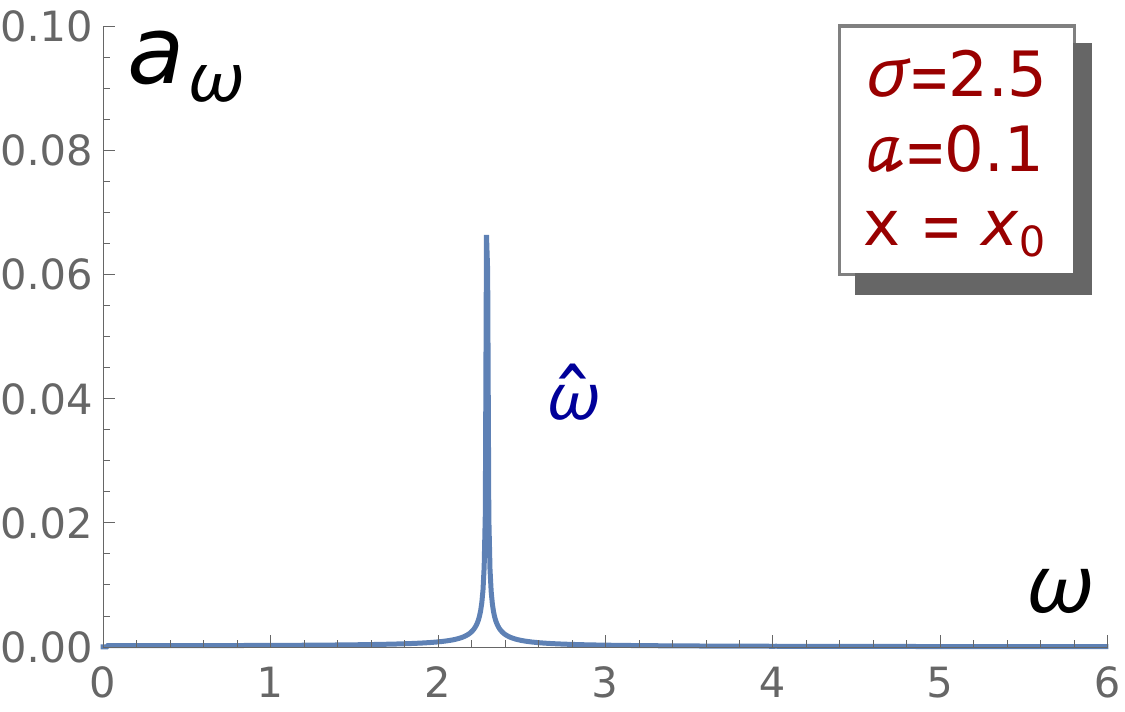}}
	\caption{\small Spectral distribution of the time series obtained by evaluating the second field component at the point $x_0$ for the evolving kink with initial orthogonal shape mode amplitude $a=0.1$. This information is shown for three representative values of model parameter: $\sigma=1.2$ (left), $\sigma=1.5$ (middle) and $\sigma=2.5$ (right).} 
	\label{fig:espectro4}
\end{figure}

\item To identify the spectral distribution of the radiation emitted in the channel of the second component of the field, the time series associated with this component must be evaluated at a point $x_B$ far from the center of the kink. Figure \ref{fig:espectro5} illustrates the results found in this case. 
The most remarkable fact in these plots is that the radiation with frequencies $\overline{\omega} + \widehat{\omega}$ (i.e. the sum of the frequencies of the two shape modes) and $3\widehat{\omega}$ is detected at the simulation  bounds. 
Note that for $\sigma=1.2$ shown in Figure \ref{fig:espectro5} (left), the most excited continuous eigenmode involves the frequency $3\widehat{\omega}$, which is the most lower than the two frequencies mentioned above. In the rest of the graphs this situation is reversed. In fact, for $\sigma=2.5$ this amplitude becomes very small.

\begin{figure}[h]
	\centerline{\includegraphics[width=0.3\textwidth]{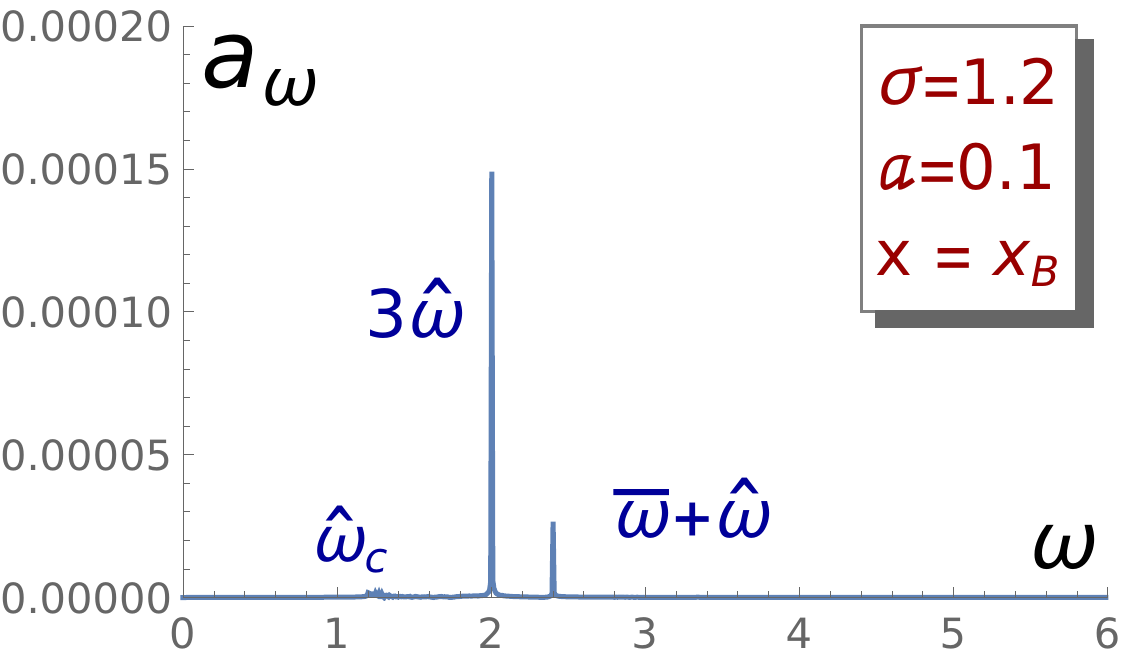} 
	\qquad
	\includegraphics[width=0.3\textwidth]{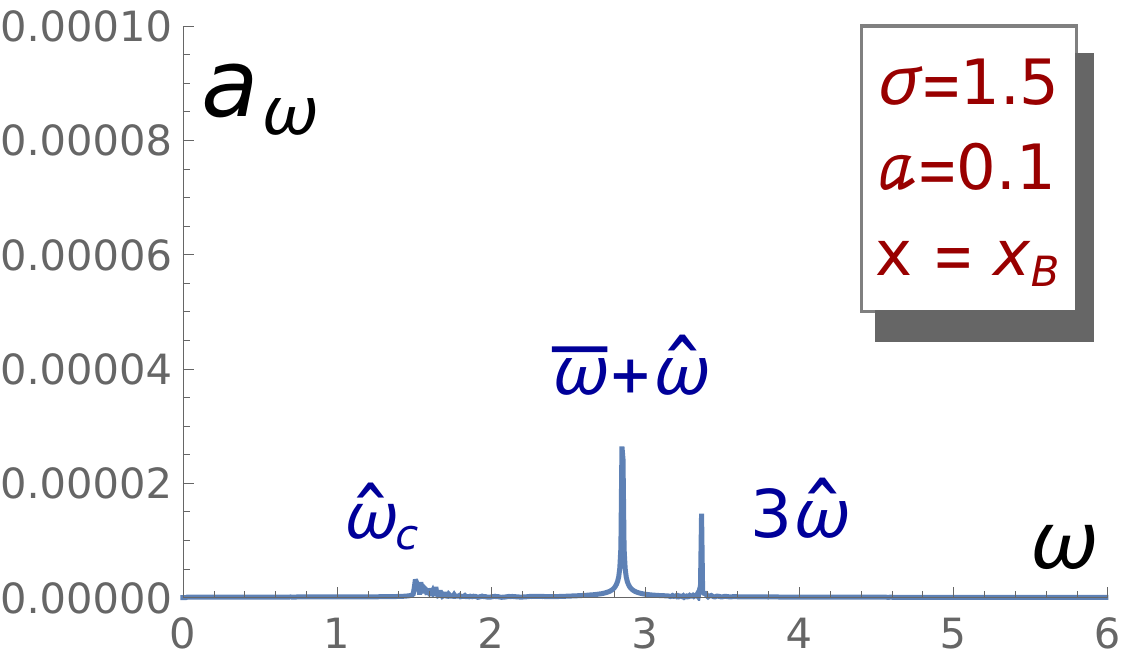} 
	\qquad
	\includegraphics[width=0.3\textwidth]{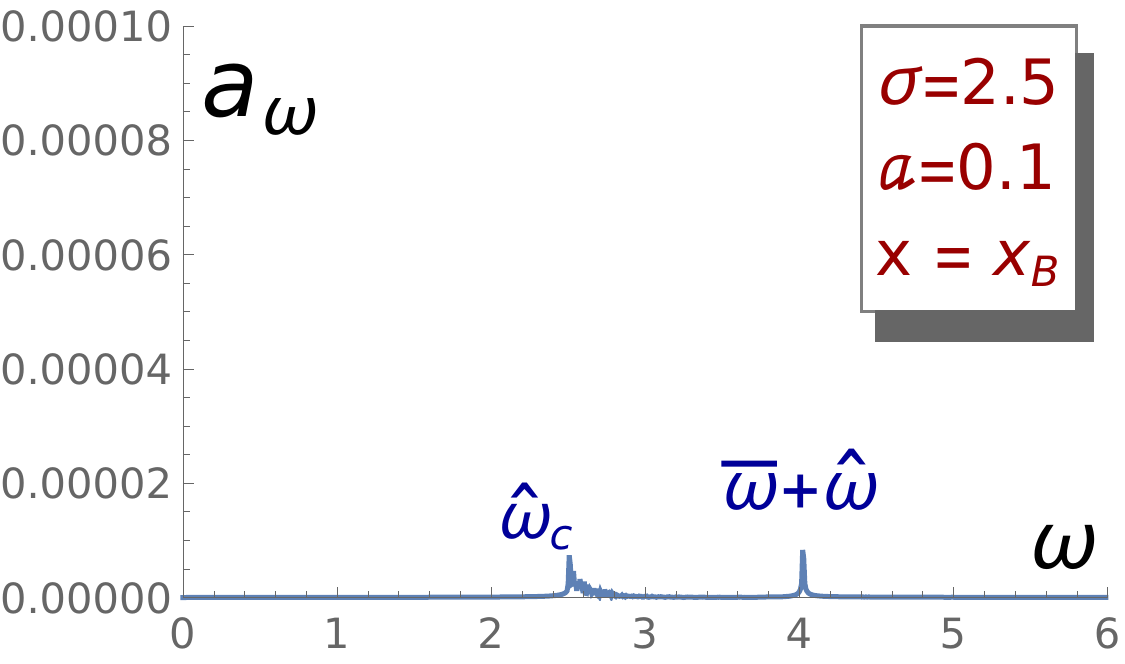}}
	\caption{\small Spectral distribution of the time series obtained by evaluating the second field component at the point $x_B$ for the evolving kink with initial orthogonal shape mode amplitude $a=0.1$. This information is shown for three representative values of model parameter: $\sigma=1.2$ (left), $\sigma=1.5$ (middle) and $\sigma=2.5$ (right).} 
	\label{fig:espectro5}
\end{figure}
\end{itemize}

\begin{figure}[htb]
	\centerline{\includegraphics[width=0.44\textwidth]{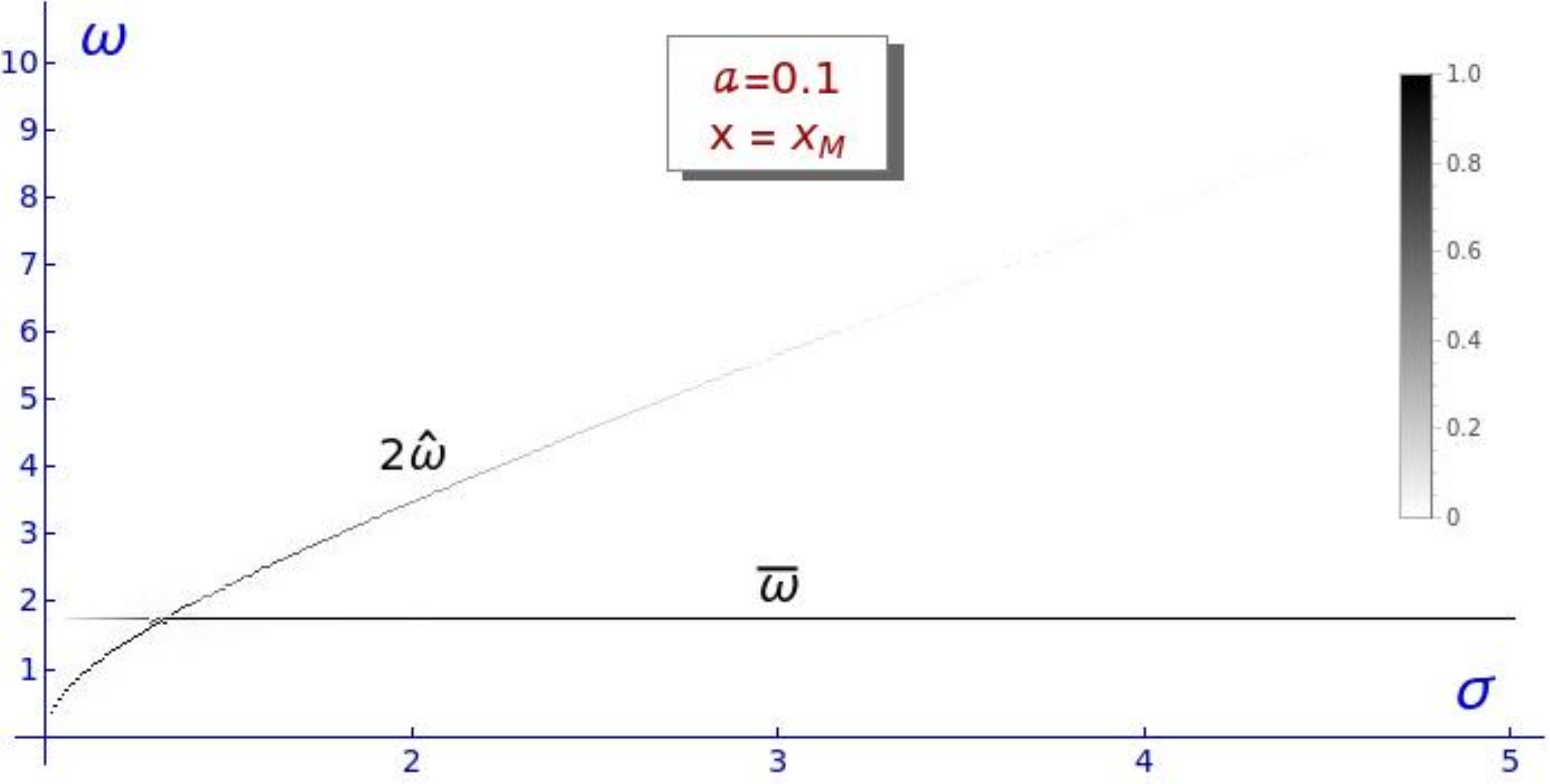} \qquad\qquad \includegraphics[width=0.44\textwidth]{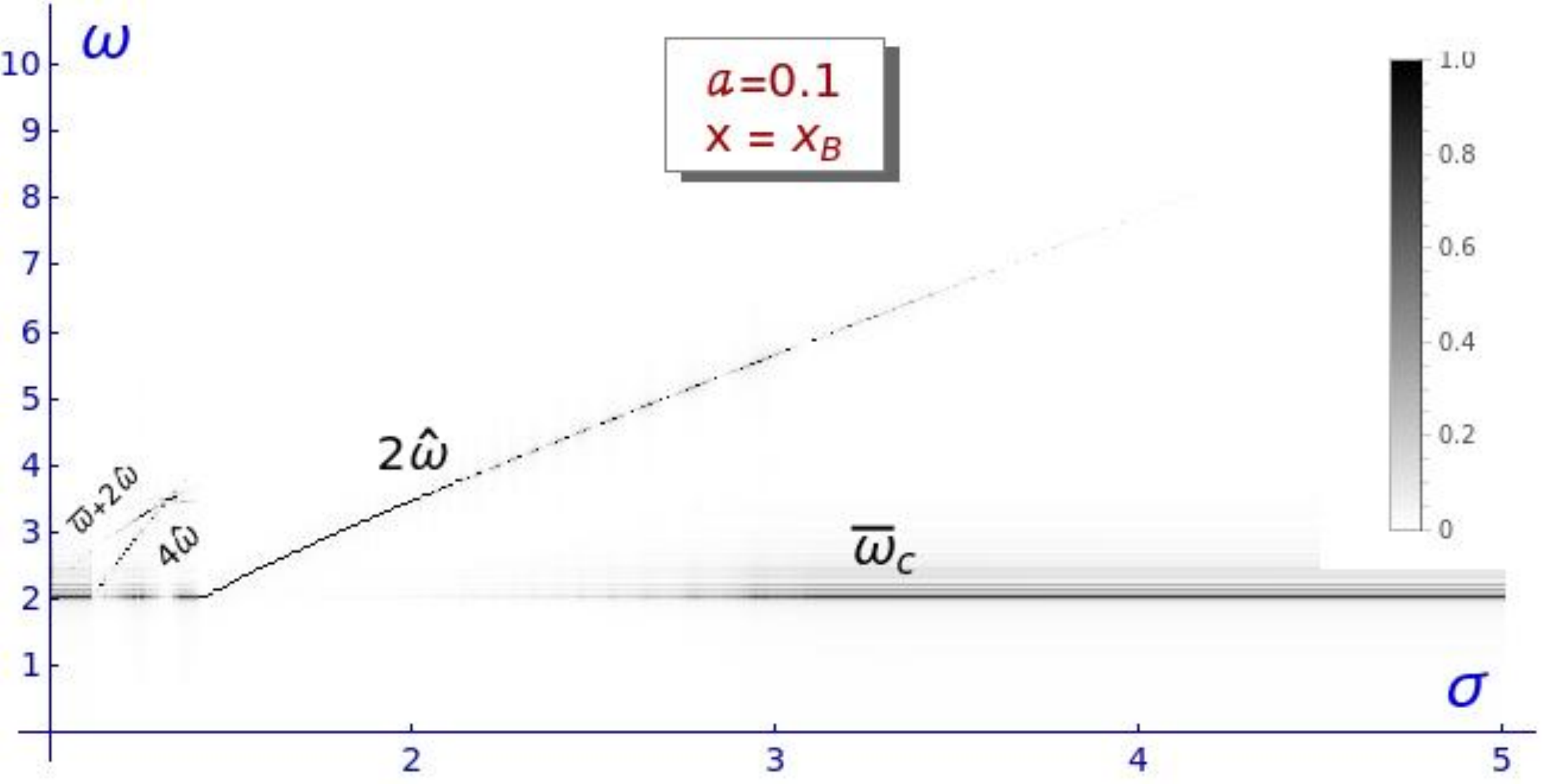}} 
	\caption{\small Frequencies of the excited longitudinal eigenmodes at $x=x_M$ (left) and $x=x_B$ (right) when the static kink is initially excited by the orthogonal shape mode with amplitude $a=0.1$.} \label{fig:espectro8a}
\end{figure}

\begin{figure}[htb]
	\centerline{\includegraphics[width=0.44\textwidth]{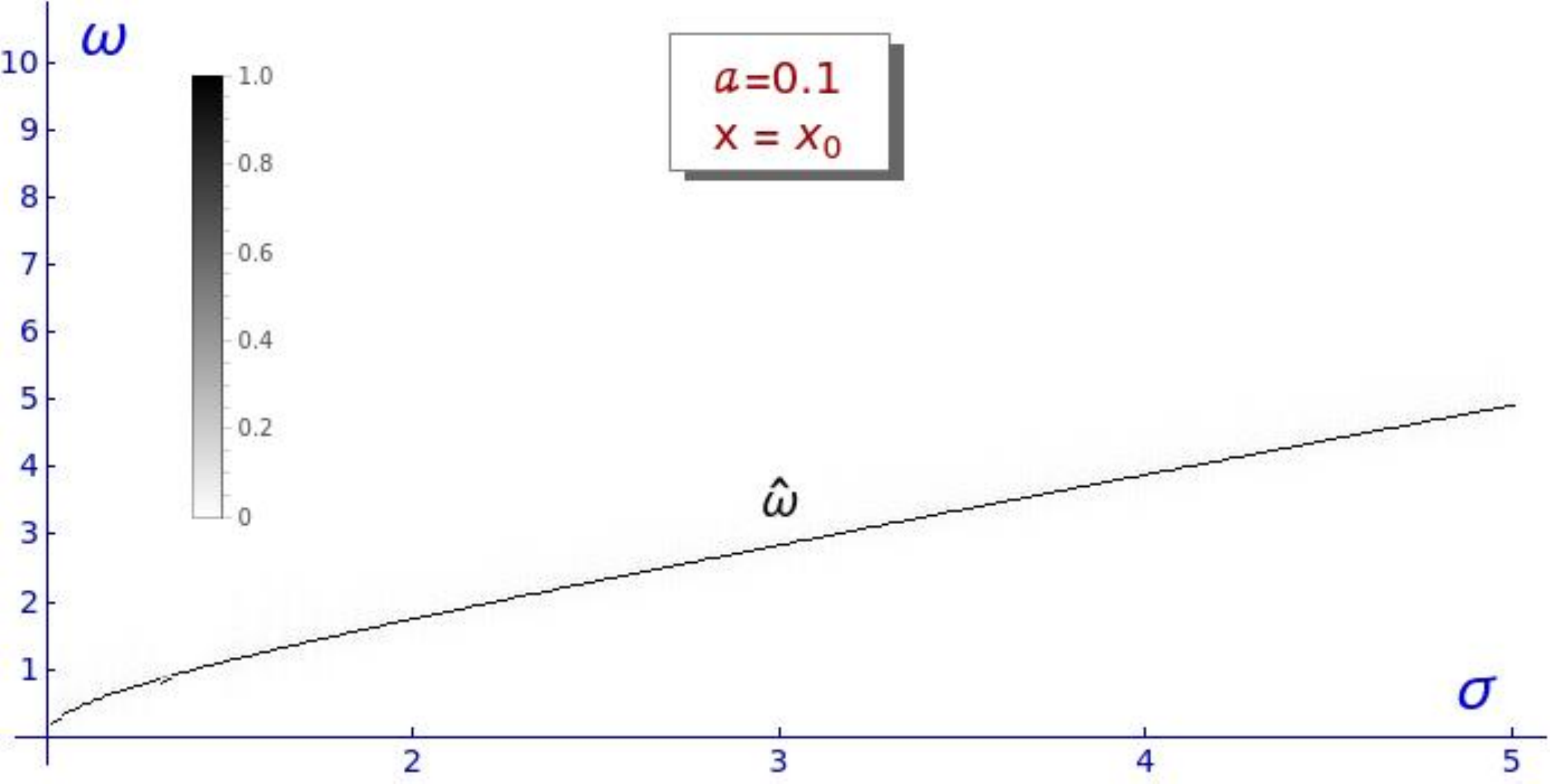} \qquad \qquad\includegraphics[width=0.44\textwidth]{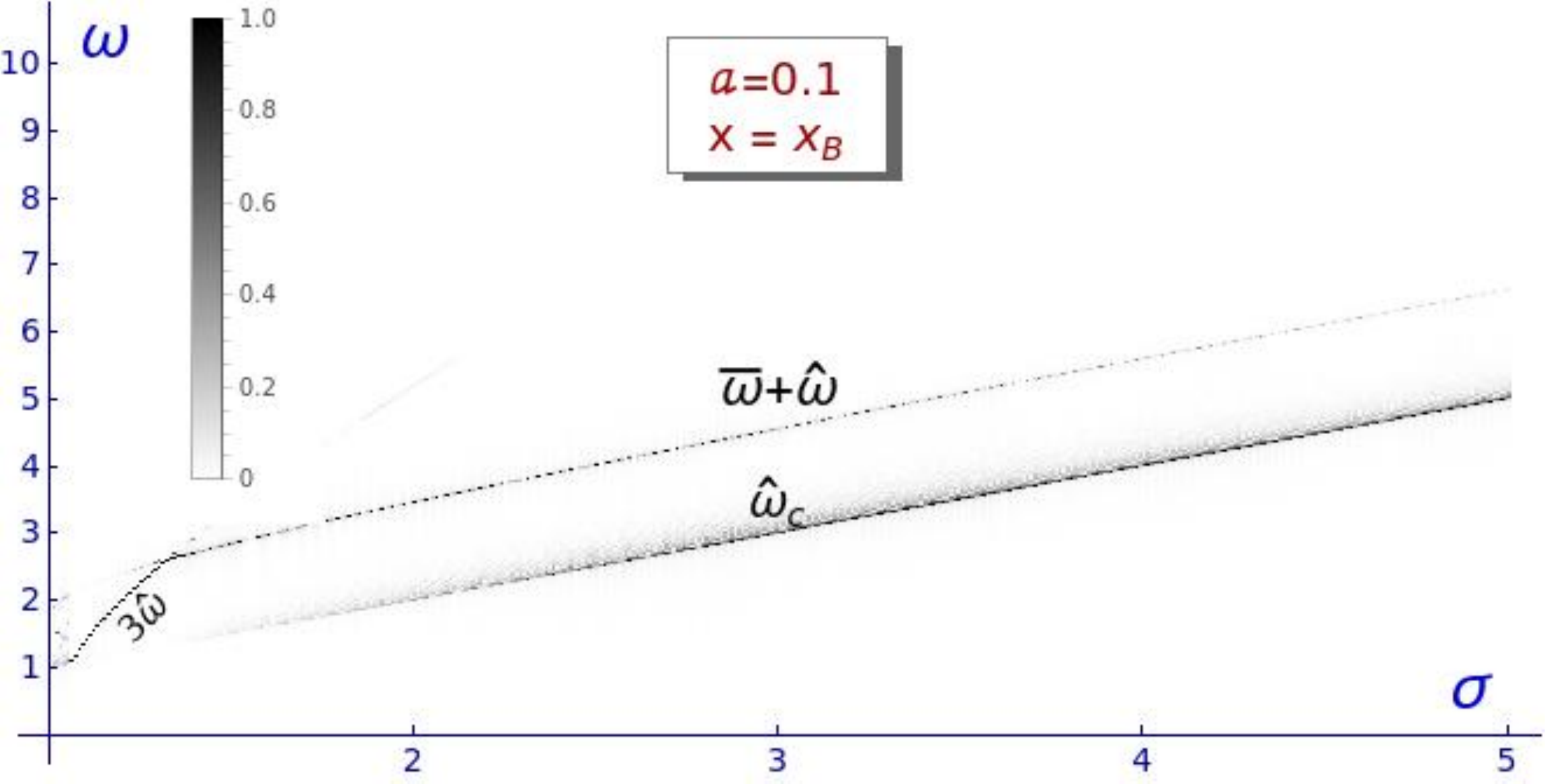}}
	\caption{\small Frequencies of the excited orthogonal eigenmodes at $x=x_0$ (left) and $x=x_B$ (right) when the static kink is initially excited by the orthogonal shape mode with amplitude $a=0.1$.} \label{fig:espectro8b}
\end{figure}

In the previous figures, the spectral information of our problem has been described for three specific values of the model parameter $\sigma$. In order to have a global understanding of the excitation process of the different eigenmodes, the previous spectral analysis has been carried out for values of $\sigma$ in the interval $\sigma\in [1,5]$ with a parameter step $\Delta \sigma = 0.01$. As before, the initial amplitude of the orthogonal shape mode in these simulations has been chosen as $a=0.1$. The results have been summarized in Figures \ref{fig:espectro8a} and \ref{fig:espectro8b}. 

The graphs show the set of frequencies that are excited as a function of the model parameter $\sigma$. These frequencies are represented by a pixel with a grayscale  pattern, where black is assigned to the most excited frequency for each $\sigma$. For the rest of frequencies (less excited)  the intensity of gray  represents the ratio between the amplitude associated with this frequency and that corresponding to the   predominant frequency mentioned above. 
If a frequency involves negligible amplitude, it is represented by a white pixel. We can observed the following characteristics:
\begin{itemize}
	\item Figure \ref{fig:espectro8a} (left) shows that $2\widehat{\omega}$ and $\overline{\omega}$ are the frequencies of the excited longitudinal eigenmodes at the point $x_M$. We can see that if $\sigma\in (1,\sigma_2)$ then the kink vibrates predominantly  with the frequency $2\widehat{\omega}$ but for $\sigma > \sigma_2$ the wobbling frequency $\overline{\omega}$ is dominant. This behavior seems to favor the less energetic eigenmode. 

	\item Figure \ref{fig:espectro8b} (left) confirms that the orthogonal shape mode is the only excited mode of the evolving kink when evaluated at the origin in the channel of the second field component. Remember that the origin can be considered as the emission source of the radiation and therefore no frequencies in the orthogonal continuous spectrum are found in the spectral analysis for this point. 

	\item Figure \ref{fig:espectro8a} (right) characterizes the frequencies of the radiation moving away from the simulation through the first field component. For $\sigma > \sigma_1$, the wobbling kink emits radiation with a frequency $2\widehat{\omega}$, though when it becomes too energetic, the radiation frequency changes to the longitudinal threshold value $\overline{\omega}_c$. 
	This happens approximately for $\sigma>3$. For $\sigma<\sigma_1$, the previous frequency $2\widehat{\omega}$ is not included in the continuous spectrum. In this regime, the radiation can propagate with frequencies close to the threshold value $\overline{\omega}_c$ but also with frequencies $4\widehat{\omega}$ or $\overline{\omega}+2\widehat{\omega}$. This behavior appears to be a consequence of the interaction of the higher order shape modes.
	
	\item Finally, Figure \ref{fig:espectro8b} (right) shows that if $\sigma>\sigma_2$ the radiation propagating in the channel of the second field component  has a dominant frequency $\overline{\omega}+\widehat{\omega}$ which is progressively replaced by values close  to the threshold value $\widehat{\omega}_c$ when the previous one is high enough. On the other hand, for $\sigma < \sigma_2$ the vibrating kink emits radiation with frequency $3\widehat{\omega}$ which in this regimen is easier to excite because it is less energetic than for higher values of $\sigma$. 

\end{itemize}

\begin{figure}[h]
	\centerline{\includegraphics[width=0.38\textwidth]{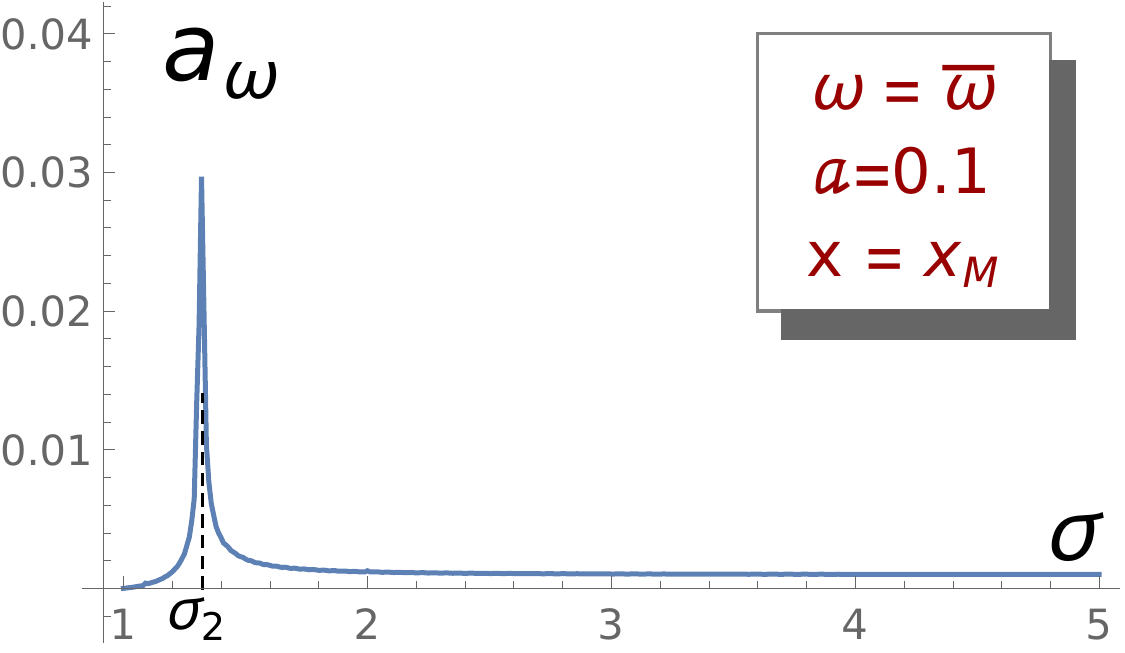}
	\qquad\qquad
	\includegraphics[width=0.4\textwidth]{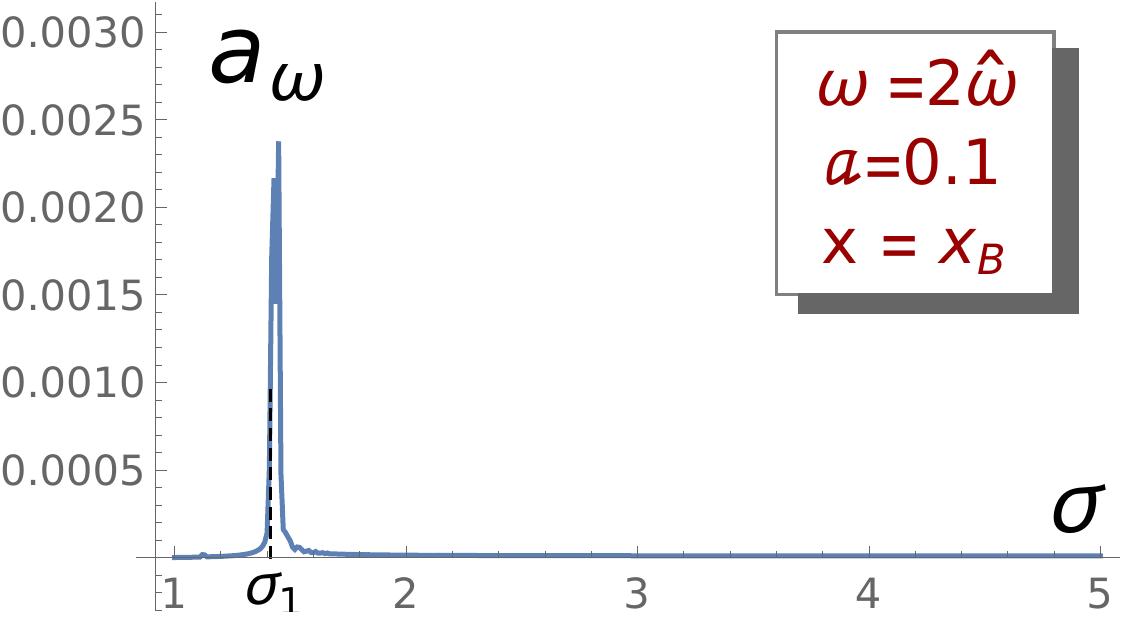} }
	\caption{\small Amplitudes of longitudinal shape mode, evaluated at $x_M$ (left), and of the longitudinal continuous mode with frequency $2\widehat{\omega}$, evaluated at $x_B$ (right), as a function of the model parameter $\sigma$.} \label{fig:espectro6}
\end{figure}

Figures \ref{fig:espectro8a} and \ref{fig:espectro8b} show that only a few eigenmodes are excited in the evolution of a kink initially excited by the orthogonal shape mode. It is also interesting to investigate the dependence of the amplitude of these eigenmodes on the value of the parameter $\sigma$. 
Figure \ref{fig:espectro6} (left) represents this dependence for the longitudinal shape eigenmodes (evaluated at the point $x_M$) for the usual value $a=0.1$. 
Note that for small values of $\sigma$ the amplitude $a_{\overline{\omega}}$ is almost insignificant but for $\sigma \approx 1.32$, when the frequency $2\widehat{\omega}$ coincides with the frequency $\overline{\omega}$ of the longitudinal shape mode, the amplitude exhibits a peak with amplitude $a_{\overline{\omega}} \approx 0.03$. Finally, the amplitude goes asymptotically to the constant value $a_\omega= 0.001$.
On the other hand, the amplitude of the longitudinal radiation mode with frequency $2\widehat{\omega}$ that escapes from the simulation is maximized for the value $\sigma \approx 1.44$ when this frequency crosses the threshold value $\overline{\omega}_c$ associated with the continuous longitudinal  spectrum. At this point, the amplitude reaches approximately the value $0.0024$, see Figure \ref{fig:espectro6} (right).

\begin{figure}[htb]
	\centerline{\includegraphics[width=0.3\textwidth]{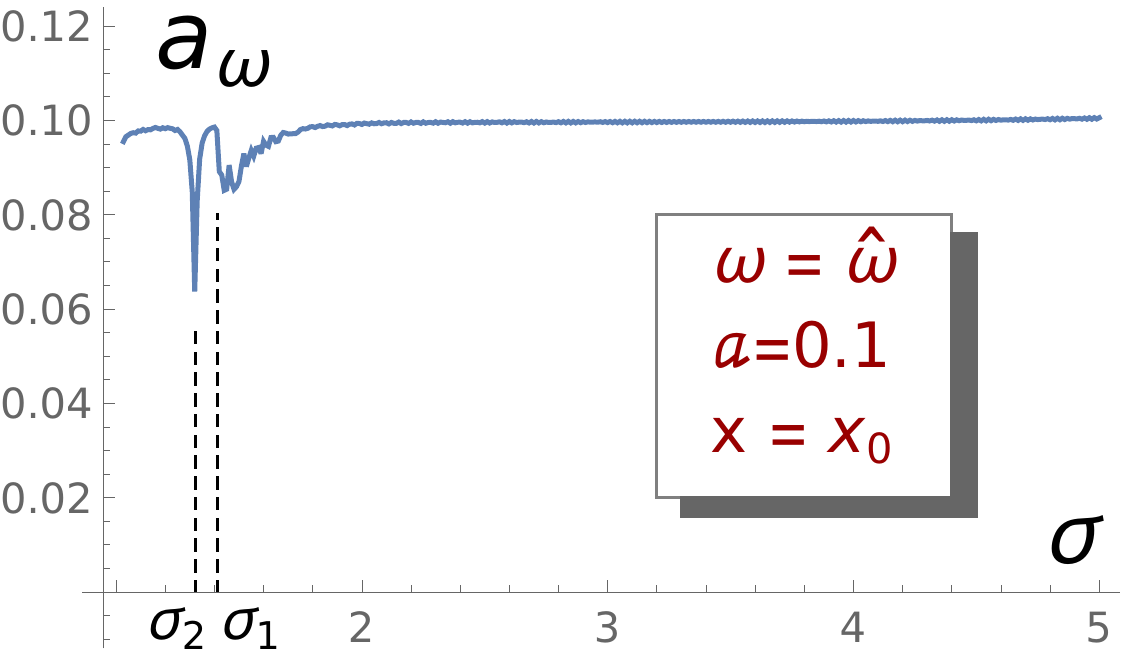} 
	\qquad
	\includegraphics[width=0.32\textwidth]{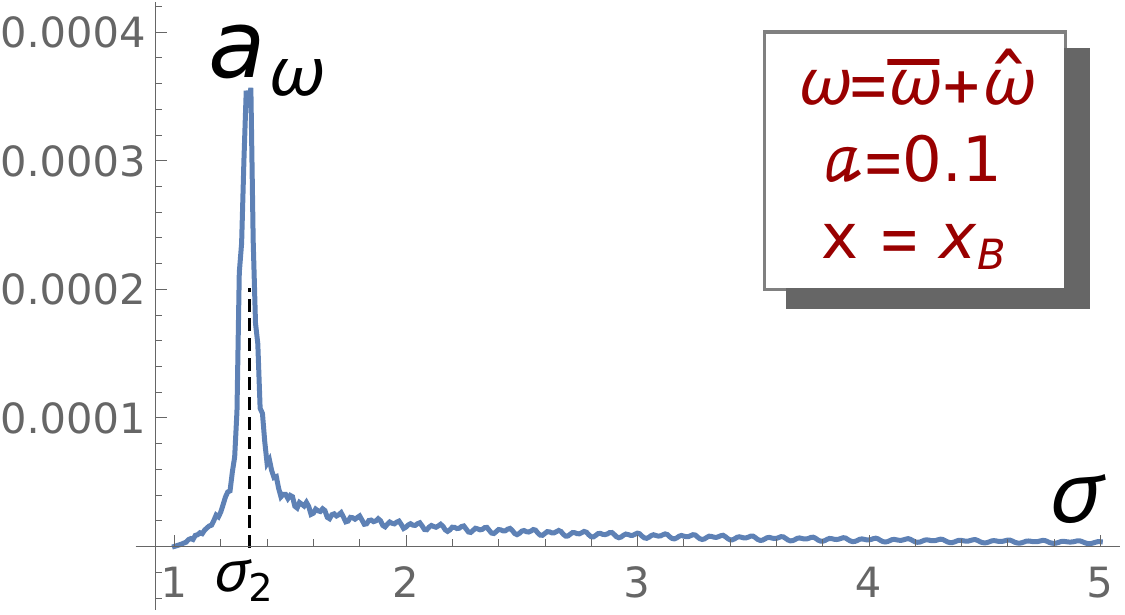}
	\qquad
	\includegraphics[width=0.32\textwidth]{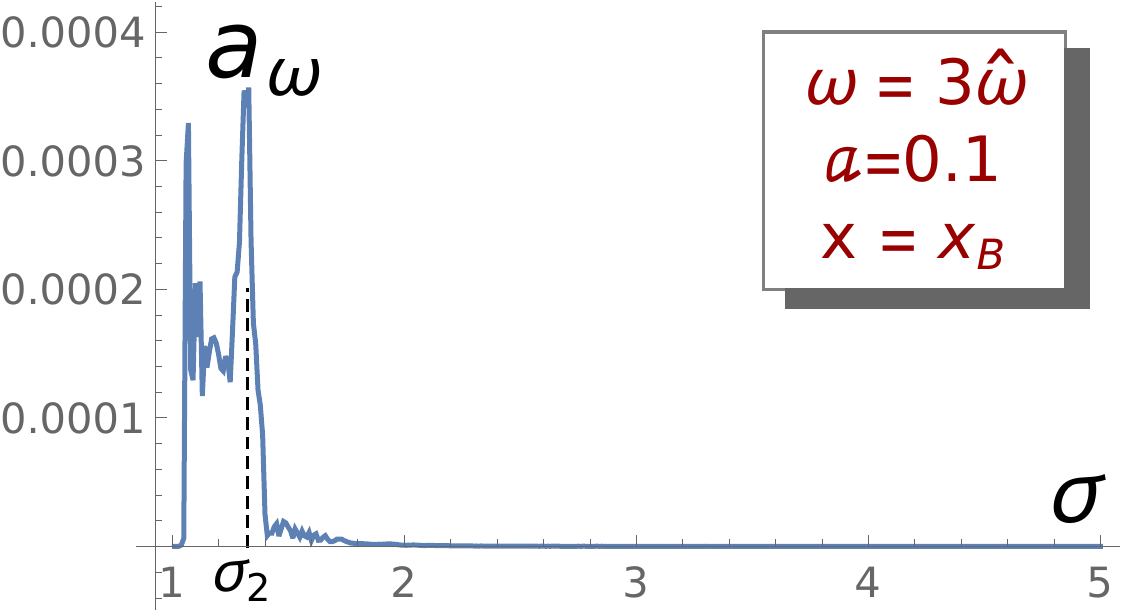}  }
	\caption{\small Amplitudes of orthogonal shape mode evaluated at $x_0$ (left) and of the orthogonal continuous mode with frequency $\overline{\omega}+\widehat{\omega}$ evaluated at $x_B$ (right) as a function of the model parameter $\sigma$.} \label{fig:espectro7}
\end{figure}

Figure \ref{fig:espectro7} (left) represents the amplitude $a_{\widehat{\omega}}$ of the orthogonal shape mode as a function of the model parameter $\sigma$. 
Here, the amplitude is almost constant $a=0.1$ although some depressions can be identified around the previously mentioned values $\sigma=1.32$ and $\sigma = 1.46$, where the amplitude of other frequencies is maximized. 
Figure \ref{fig:espectro7} (center) represents the amplitude of the orthogonal radiation eigenmodes with frequency $\overline{\omega} + \widehat{\omega}$. 
The peak of this graph arises at the value $\sigma=1.32$. 
Finally, Figure \ref{fig:espectro7} (right) represents the amplitude of the eigenmode with frequency $3\widehat{\omega}$. 
Here, we find that this eigenmode is highly excited for $\sigma < 1.36$, showing two peaks located at $\sigma= 1.068$ and $\sigma=1.32$. After that, the amplitude decreases.

In the next two section, perturbation expansion theory will be used to explain some of the surprising results found in the present section. In particular, we will apply the techniques of Manton \& Merabet \cite{Manton1997} and   Barashenkov \& Oxtoby \cite{Barashenkov2009} to our problem. Obviously, some modifications of these two approaches will be implemented because we are dealing with a two-component field theory model with a particular initial setting. The results found in each case will be compared.
In particular, we are very interested in understanding the suppression mechanism of longitudinal radiation emission with frequency $2\overline{\omega}$. We remember that in the $\phi^4$-model radiation with this frequency was emitted, however in our extended model it is replaced by longitudinal radiation with frequency $2\widehat{\omega}$. 
Another feature which will be examined is the presence of orthogonal radiation with frequency $\overline{\omega}+ \widehat{\omega}$ and, more surprisingly, with frequency $3\widehat{\omega}$ at the same scale. 
This means that the radiation emission  in the latter case is not a  higher order  effect 
in the perturbation expansion, but must be explained in the same order as the previous frequency. The explanation of this behavior lies in the interaction between the longitudinal and orthogonal shape modes in the problem presented in the present section.

\section{Perturbative approach: Manton and Merabet approach}\label{MantonMerabet}

As previously mentioned, in \cite{Manton1997} Manton and Merabet applied a perturbation expansion to the $\phi^4$-model to explain the emission of radiation by a wobbling kink with twice its natural vibration frequency, and they  also estimate the decay rate of the wobbling amplitude due to this process. In the sequel we will follow the essence of this method to analyze the system we are dealing with.

In this case, the components of the field \eqref{pde1}--\eqref{pde2} will be written as
\begin{equation}
\begin{array}{l}
\phi(x,t) = \phi_K(x) + \overline{a}(t) \, \overline{\eta}_D(x) + \overline{\eta}(x,t) , \\ [1ex]
\psi(x,t) =  \phantom{\phi_K(x) +\ } \widehat{a}(t)  \, \widehat{\eta}_D (x) + \widehat{\eta}(x,t),  
\end{array}
\label{MantonExpansion}
\end{equation}
where $\phi_K(x)$, $\overline{\eta}_D(x)$ and $\widehat{\eta}_D (x)$ where defined respectively in (\ref{kinkdef}), (\ref{shapemode1}) and (\ref{shapemode2}). 
The time dependent functions $\overline{a}(t)$ and $\widehat{a}(t)$ respectively describe the evolution of the amplitude of the longitudinal and orthogonal shape modes. 
On the other hand, the space and time dependent functions $\overline{\eta}(x,t)$ and $\widehat{\eta}(x,t) $ determine the evolution of the longitudinal and orthogonal radiation eigenmodes. 
For the sake of clarity, from now on the variables on which both the functions $\overline{a}$ and $\widehat{a}$ and the eigenfunctions $\overline{\eta}_D$, $\widehat{\eta}_D$, $\overline{\eta }$, and $\widehat{\eta}$ depend, are not written explicitly.
Thus, if we substitute the expression (\ref{MantonExpansion}) into the field equations (\ref{pde1}) and (\ref{pde2}) we find 
\begin{eqnarray}
	&&\hspace{-0.7cm} \overline{\eta}_D  (3 \, \overline{a} + \overline{a}_{tt}) + \overline{\eta}_{tt} - \overline{\eta}_{xx} + \overline{\eta}  \, (6 \phi_K^2(x)-2) + 6\, \overline{a}^2 \, \overline{\eta}_D^2 \, (\overline{\eta}+\Phi_K(x))  \nonumber \\ 
	&& + 6 \, \overline{a} \, \overline{\eta} \, \overline{\eta}_D \,  (\overline{\eta}+2 \phi_K(x)) +2 \overline{a}^3 \, \overline{\eta}_D^3 + 6 \, \overline{\eta}^2  \, \phi_K(x) + 2\, \overline{\eta}^3 + 2 \, \overline{\eta}\, \widehat{\eta}^2  \label{ecu1total} \\ 
	&& + 2 \, \overline{a} \, \widehat{\eta}^2 \, \eta_D + 4\, \widehat{a} \, \overline{\eta}\,\widehat{\eta} \, \widehat{\eta}_D + 4 \, \overline{a}  \, \widehat{a}  \, \widehat{\eta} \,  \overline{\eta}_D  \, \widehat{\eta}_D + 2 \,    \widehat{a}^2\, \overline{\eta}\, \widehat{\eta}_D^2  \nonumber  \\ 
	&& +2 \, \overline{a}  \, \widehat{a}^2 \, \overline{\eta}_D  \, \widehat{\eta}^2 + 2 \, \widehat{\eta}^2  \, \phi_K(x) + 4\,\widehat{a}  \,  \widehat{\eta} \,  \widehat{\eta}_D \, \phi_K(x)+ 2 \, \widehat{a}^2  \, \widehat{\eta}_D^2  \, \phi_K(x) =0 \nonumber 
\end{eqnarray}
for the first component, and 
\begin{eqnarray}
	&& \hspace{-0.7cm}  \widehat{\eta}_D  \, (\widehat{a}_{tt}+(\sigma^2-1) \, \widehat{a}) + (\sigma^2-2)   \,  \widehat{\eta} + 2 \, \overline{\eta}^2 \, \widehat{\eta} + 2  \widehat{\eta}^3 + 4 \, \overline{a}\, \overline{\eta}\, \widehat{\eta} \, \overline{\eta}_D  \nonumber  \\
	&& + 2\,\overline{a}^2 \,  \widehat{\eta} \, \overline{\eta}_D^2  + 2\, \widehat{a} \, \overline{\eta}^2 \, \widehat{\eta}_D + 6 \, \widehat{a}   \, \widehat{\eta}^2  \, \widehat{\eta}_D + 4\, \overline{a}\, \widehat{a}\, \overline{\eta}\,\overline{\eta}_D \, \widehat{\eta}_D  \label{ecu2total} \\ 
	&& + 2 \, \overline{a}^2 \widehat{a} \, \overline{\eta}_D^2 \, \widehat{\eta}_D + 6\, \widehat{a}^2\,  \widehat{\eta} \, \widehat{\eta}_D^2 + 2\, \widehat{a}^3  \, \widehat{\eta}_D^3 + 4\, \overline{\eta}\,  \widehat{\eta} \, \phi_K(x)  \nonumber \\
	&& + 4\, \overline{a}  \,  \widehat{\eta}  \, \overline{\eta}_D  \, \phi_K(x) + 4 \, \widehat{a}  \, \overline{\eta} \, \widehat{\eta}_D\,  \phi_K(x) + 4\, \overline{a}  \,  \widehat{a}  \,  \overline{\eta}_D \, \widehat{\eta}_D\, \phi_K(x) + 2 \,   \widehat{\eta}  \, \phi_K^2(x)   +  \widehat{\eta}_{tt} -  \widehat{\eta}_{xx} = 0\nonumber 
\end{eqnarray}
for the second one. In the problem at hand, we assume that the amplitude of the orthogonal shape mode is small, so  we can neglect terms of the form $a^3$, $\eta^2$, $\eta a$, etc. in (\ref{ecu1total}) and (\ref{ecu2total}), where $a$ stands for either $\overline{a}$ or $\widehat{a}$ and $\eta$ stands for $\overline{\eta}$ or $ \widehat{\eta}$. 
Furthermore, the amplitudes of the emitted radiation are assumed to be much smaller than those associated with the shape modes. Under these assumptions the equations (\ref{ecu1total}) and (\ref{ecu2total}) reduce to
\begin{equation}
	\overline{\eta}_D (3 \, \overline{a} + \overline{a}_{tt}) + \overline{\eta}_{tt} - \eta_{xx} + \overline{\eta} ( 6 \, \phi_K^2(x)-2 ) + 6 \overline{a}^2 \, \overline{\eta}_D^2 \phi_K(x) + 2 \, \widehat{a}^2 \, \widehat{\eta}_D^2 \, \phi_K(x) \approx 0 \label{ecu1despre}
\end{equation}
and 
\begin{equation}
	 \widehat{\eta}_D \,  ( \widehat{a}_{tt} + (\sigma^2 -1) \, \widehat{a}) +  \widehat{\eta}_{tt} -  \widehat{\eta}_{xx}+(-2+\sigma^2 + 2 \phi_K^2(x))  \, \widehat{\eta} + 4 \overline{a}\, \widehat{a} \, \overline{\eta}_D \, \widehat{\eta}_D \, \phi_K(x)  \approx 0 . \label{ecu2despre}
\end{equation}
The projection of (\ref{ecu1despre}) onto the longitudinal shape mode $\overline{\eta}_D$ leads to the relation 
\begin{equation}
\overline{a}_{tt} + 3 \, \overline{a} + \frac{9}{16}\,\pi\, \overline{a}^2 + \frac{3}{8} \, \pi \, \widehat{a}^2 = 0, \label{amplilong}
\end{equation}
while the projection of (\ref{ecu2despre}) onto the orthogonal shape mode $ \widehat{\eta}_D$ yields 
\begin{equation}
\widehat{a}_{tt} + (\sigma^2 -1)  \, \widehat{a} + \frac{1}{4} \, \pi \, \overline{a} \, \widehat{a} = 0 .\label{ampliorto}
\end{equation}
On the other hand, substituting \eqref{amplilong} in \eqref{ecu1despre} we get that the equation for the longitudinal radiation component $\overline{\eta}$ is
\begin{equation}
	\overline{\eta}_{tt} -\overline{\eta}_{xx} + (6 \phi_K^2(x)-2 )  \, \overline{\eta} + \Big[6 \, \overline{\eta}_D^2 \, \phi_K(x) - \frac{9}{16}\, \pi\, \overline{\eta}_D \Big] \overline{a}^2 + \Big[ 2 \widehat{\eta}_D^2  \, \phi_K(x) - \frac{3}{8}\, \pi \, \overline{\eta}_D \Big] \widehat{a}^2 = 0, \label{radilong}
\end{equation}
while substituting \eqref{ampliorto} in \eqref{ecu2despre} we get for the orthogonal component $\widehat{\eta}$ the equation
\begin{equation}
 \widehat{\eta}_{tt} -  \widehat{\eta}_{xx} + (\sigma^2- 2 + 2 \phi_K^2(x))   \, \widehat{\eta} + \Big[ 4\,  \overline{\eta}_D\, \phi_K(x) - \frac{\pi}{4} \Big] \,\widehat{\eta}_D  \, \overline{a} \, \widehat{a} = 0 \label{radiorto}.
\end{equation}
The problem addressed in this work is characterized by the excitation of the orthogonal shape mode while the longitudinal  remains initially unexcited. We are interested in analyzing the mechanism of energy transfer between the shape modes and the emission of radiation  in this process. 
It is clear that the quadratic term $\widehat{a}^2$ arising in (\ref{amplilong}) implies the subsequent excitation of the longitudinal shape mode since  the evolution of $\overline{a}$ is forced by this term. Therefore, we can assume that at first-order the amplitude $\widehat{a}$ of the orthogonal shape mode  follows the expression
\begin{equation}
	\widehat{a} (t) = \widehat{a}_0 \,  \sin (\widehat{\omega} \, t) = \widehat{a}_0  \, \sin (\sqrt{\sigma^2 -1 } \,\, t ).
	\label{ampliorto2}
\end{equation}
If we substitute (\ref{ampliorto2}) into (\ref{amplilong}) and the initial conditions $\overline{a}(0) = {\overline{a}_t}(0)=0$ are implemented, then the evolution of the amplitude $\overline{a}(t)$ of the longitudinal mode is given by
\begin{equation}
\overline{a}(t) = \frac{\pi\, \widehat{a}_0^2 }{16} \Big[ -1 + \frac{4(\sigma^2-1)}{4\sigma^2- 7} \cos (\sqrt{3} t) - \frac{3}{4\sigma^2-7} \cos (2\sqrt{\sigma^2-1}\,t)\,  \Big] .   
\label{amplilong2}
\end{equation}
We can see that the above formula anticipates that $\sigma=\sigma_1=\sqrt{7}/2$ is special  because for precisely that value the denominators in (\ref{amplilong2}) cancel. 
We can check in Figures \ref{fig:espectro6} and \ref{fig:espectro7} that for this value of the coupling constant $\sigma$ the amplitude of the longitudinal shape mode exhibits an abrupt peak. 
In this case a resonance arises between the frequencies $\overline{\omega}$ and $2\widehat{\omega}$. However, in this article we will not deal with this particular case because we are interested in the more general interval $\sigma > \sigma_1=\sqrt{2}$ where the value $2\widehat{\omega}$ enters into resonance with the continuous frequencies in the longitudinal channel. 

From (\ref{amplilong2}) it is clear that $\overline{a} \sim O(\widehat{a}^2)$, ie the scales of the amplitudes of the shape modes are different. Taking into account that
\[
(\widehat{a} (t))^2 = \frac{1}{2}  \, \widehat{a}_0^2 \left( 1-\cos (2 \, \widehat{\omega} \,  t)\right) = \frac{1}{2} \,  \widehat{a}_0^2 \left(1-\cos (2\sqrt{\sigma^2-1} \, t) \right)
\]
and that the response of $\overline{\eta}$ to the time-independent source is itself time-independent and carries no energy, then the non-trivial evolution of longitudinal radiation is governed by the equation 
\begin{equation}
\overline{\eta}_{tt} - \overline{\eta}_{xx} + (6 \phi_K^2(x)-2)  \, \overline{\eta} = f(x) \, e^{i \, 2 \widehat{\omega} t} , \label{radilong3}
\end{equation}
which has been derived from the second-order truncation of (\ref{radilong}) in $ \widehat{a}_0$. This  means that  terms involving  $\overline{a}^2(t)$ in (\ref{radilong}) have been ignored because $\overline{a} \sim O(\widehat{a}^2)$, as  mentioned above. 
At this point it is necessary to clarify that, in order to simplify the tedious calculations that must be carried out, we have decided to use imaginary exponentials in (\ref{radilong3}). But in fact, at the end of the calculation, only the real part of the solutions of (\ref{radilong3}) will be relevant, so this choice does not affect the final results of this section. In fact, the function $f(x)$ of (\ref{radilong3}) has the following expression:
\begin{equation}
	f(x) = \frac{1}{2}  \, \widehat{a}_0^2\, \Big(2\,  \widehat{\eta}_D^2 \, \phi_K(x) - \frac{3}{8}  \, \pi  \, \overline{\eta}_D \Big)  = - \frac{1}{2}  \widehat{a}_0^2 \Big( \frac{3}{8} \pi \, {\rm sech} \, x \tanh x - 2 \, {\rm sech}^2 x \tanh x \Big) \, .\label{funcionfdex}
\end{equation}
In these circumstances it can be assumed that the function describing the longitudinal radiation has the form
\begin{equation}
\overline{\eta}(x,t) =  \overline{\eta}(x) \, e^{i \,2 \widehat{\omega} t},
 \label{etaxt}
\end{equation}
which leads to the following ordinary differential equation in coordinate space:
\begin{equation}
- \overline{\eta}\,''(x) + [-2+6\, \phi_K^2(x)-4\widehat{\omega}^2 ] \, \overline{\eta}(x) = f(x) .\label{radilong4}
\end{equation}
A similar argument is used to obtain an ordinary differential equation for the orthogonal component. The expression (\ref{radiorto}) contains the product of the amplitudes 
\begin{eqnarray}
 \overline{a}(t) \widehat{a}(t)  \!\!&\!\!=\!\!&\!\! \frac{\pi\, \widehat{a}_0^3 \, (17-8 \sigma^2)}{32 \left(4 \sigma^2-7\right)} \sin (\widehat{\omega} t) + \frac{\pi  \,  \widehat{a}_0^3 \!\left(\sigma^2-1\right) }{8 \left(4
 	\sigma^2-7\right)}\sin [(\overline{\omega}+\widehat{\omega} ) t ] \nonumber \\  [1ex]
\!\!&\!\! \!\!&\!\! - \frac{\pi\,  \widehat{a}_0^3\! \left(\sigma^2-1\right) }{8 \left(4 \sigma^2-7\right)} \sin [ (\overline{\omega}-\widehat{\omega} ) t ] -\frac{3 \pi  \,  \widehat{a}_0^3 }{32 \left(4 \sigma^2-7\right)} \sin (3 \, \widehat{\omega} t ) ,  \label{productAs}
\end{eqnarray}
that contains sinusoidal functions with four different frequencies: 
\begin{equation}\label{frequenciesell}
\omega_1= \widehat{\omega}, \quad \omega_2=3 \, \widehat{\omega}, \quad\omega_3=\overline{\omega}+\widehat{\omega}, \quad \text{and}\quad \omega_4=\overline{\omega}-\widehat{\omega}. 
\end{equation}
Note the presence, a priori unexpected, of $\omega_2$, three times the natural vibration frequency $\widehat{\omega}$ of the orthogonal shape mode. Due to the linearity of (\ref{radiorto}), the effect of each term in (\ref{productAs}) on the radiation solution can be studied separately. Therefore, the emission of radiation with frequency $\omega_\ell$ can be examined by analyzing the following four partial differential equations in   complex  form
\begin{equation}
 \widehat{\eta}_{tt} -  \widehat{\eta}_{xx} + (\sigma^2-2+2 \phi_K^2(x))  \widehat{\eta} = g_\ell(x) e^{i\omega_\ell t} \, ,
\end{equation}
where
\begin{eqnarray*}
g_1(x) \!\!&\!\!=\!\!&\!\!  - \frac{ \widehat{a}_0^3 \,\pi(8 \sigma^2 -17) \,  (\pi-16 \,{\rm sech}\, x \tanh^2 x)\, {\rm sech}\, x}{128 (4\sigma^2 - 7)} \, ,\\
g_2(x) \!\!&\!\!=\!\!&\!\! - \frac{ \widehat{a}_0^3 \, 3\,\pi  (\pi-16 \,{\rm sech}\, x \tanh^2 x)\, {\rm sech}\, x}{128 (4\sigma^2 - 7)}  \, , \\
g_3(x) = - g_4(x) \!\!&\!\!=\!\!&\!\!  \frac{ \widehat{a}_0^3 \,\pi(\sigma^2 -1)  (\pi-16 \,{\rm sech}\, x \tanh^2 x) \, {\rm sech}\, x}{32 (4\sigma^2 - 7)}   \, .
\end{eqnarray*}
As before, it is assumed that for each frequency $\omega_\ell$ this orthogonal radiation can be written as
\[
 \widehat{\eta} (x,t) =  \widehat{\eta} (x) \, e^{i\omega_\ell t}  \, ,
\]
which leads to the four ordinary differential equations
\begin{equation}
- \widehat{\eta}''(x) + (\sigma^2-2+2 \, \phi_K^2(x)-\omega_\ell^2) \,  \widehat{\eta}(x) = g_\ell(x) , \quad    \ell=1,2,3,4 .
\label{radiorto4}
\end{equation}
Now the equations (\ref{radilong4}) and (\ref{radiorto4}) will be solved to determine the radiation emission of the shape modes in each channel. As  mentioned above,  this study will be carried out in the  $\sigma>\sigma_1$ regime where twice the frequency of the orthogonal shape mode is included in the longitudinal continuous spectrum. We  explicitly recall the notation introduced in (\ref{continuous1}) for the first component of the continuous longitudinal  eigenfunctions:
\begin{equation}
\overline{\eta}_q(x) = e^{iqx} \left( -1-q^2+3\,\phi_K^2(x) -3 i q \, \phi_K(x)\right)  .   \label{eta1}
\end{equation}
From the dispersion relation, we have that $q=\sqrt{\omega^2-4}$ with $|\omega| \geq 2$. Note that the Wronskian associated with these solutions 
\[
\overline{W}_q= 
W[\overline{\eta}_q(x), \overline{\eta}_{-q}(x)] = -2 i q (q^2+1)(q^2+ 4)
\]
is constant. 
The solution of (\ref{radilong4}) that describes the radiation emitted from the origin with frequency $2 \widehat{\omega}$ is given by the expression
\begin{equation}
\overline{\eta}(x) = - \frac{\overline{\eta}_{-q_1}(x)}{\overline{W}_{q_1}}  \int_{-\infty}^x \overline{\eta}_{q_1} (\xi) f(\xi) d\xi - \frac{\overline{\eta}_{q_1}(x)}{\overline{W}_{q_1}}  \int_x^\infty \overline{\eta}_{-q_1} (\xi) f(\xi) d\xi  \, ,
\label{radilong5}
\end{equation}
where $q_1=\sqrt{(2\widehat{\omega})^2-4}=2\sqrt{\sigma^2-2}$. 
The asymptotic behavior of (\ref{radilong5}) can be identified analytically and turns out to be 
\begin{equation*}
\overline{\eta}(x) \stackrel{x\rightarrow \infty}{\longrightarrow}  e^{-iq_1x} \  \widehat{a}_0^2 \, \frac{\pi}{16} \ \frac{q_1(2+3 i q_1-q_1^2)}{(q_1^2+1)\, \sinh \frac{\pi q_1}{2}}  \, .
\end{equation*}
Multiplying the above result  by $e^{i \,2 \widehat{\omega} t}$ 
 and using  \eqref{etaxt}
 we get $\overline{\eta}(x,t)$. Then, taking only its real part, which is what really interests us, it can be determined that the behavior of the longitudinal radiation of frequency $2\widehat{\omega}$ turns out to be
\begin{equation}
\overline{\eta}(x,t) \stackrel{x\rightarrow \infty}{\longrightarrow}  \frac{\pi \sqrt{\sigma^2-2}}{4 \sinh (\pi \sqrt{\sigma^2-2})}\ \sqrt{\frac{\sigma^2-1}{4 \sigma^2-7}} \  \widehat{a}_0^2 \cos \left(2 \widehat{\omega} t -q_1  x + \delta_1 \right), \quad
\delta_1=\arctan \left( \frac{3q_1}{2-q_1^2}\right)  .
\label{radilong6}
\end{equation}
Therefore, the amplitude of the radiation emitted with frequency $2\widehat{\omega}$ in the longitudinal channel is
\begin{equation}
a_{2 \widehat{\omega}} =  \frac{\pi \sqrt{\sigma^2-2}}{4 \sinh (\pi \sqrt{\sigma^2-2})} \ \sqrt{\frac{\sigma^2-1}{4 \sigma^2-7}} \  \widehat{a}_0^2  . 
\label{radilong7}
\end{equation}
It can be checked from (\ref{radilong7}) that $a_{2 \widehat{\omega}} \sim O( \widehat{a}_0^2)$. 

\begin{figure}[htb]
	\centerline{\includegraphics[width=0.31\textwidth]{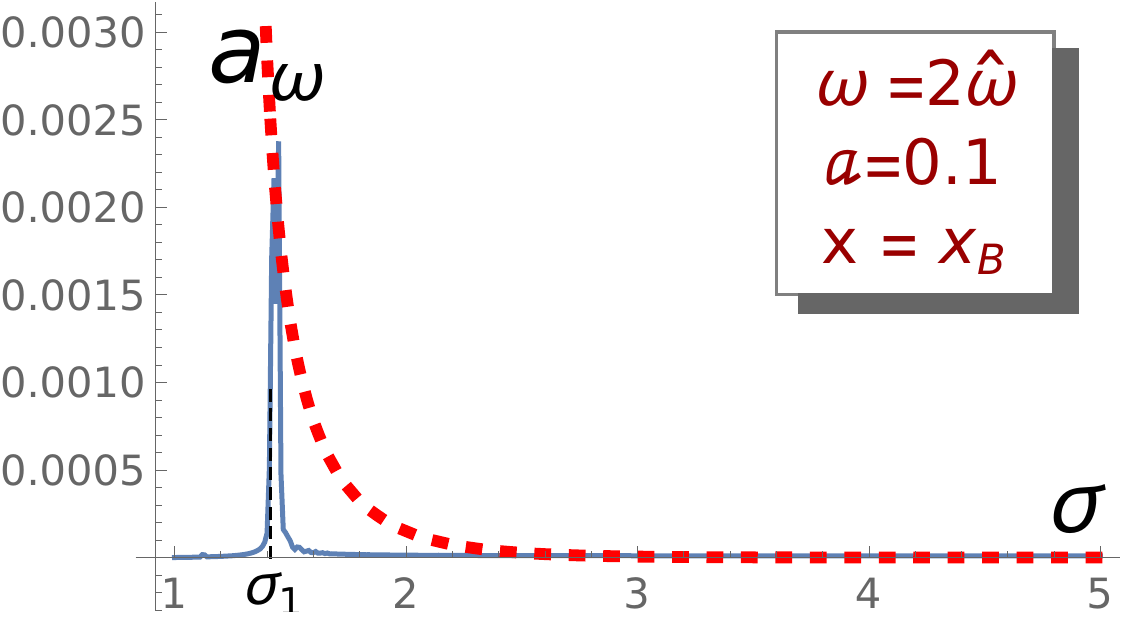} \qquad\includegraphics[width=0.31\textwidth]{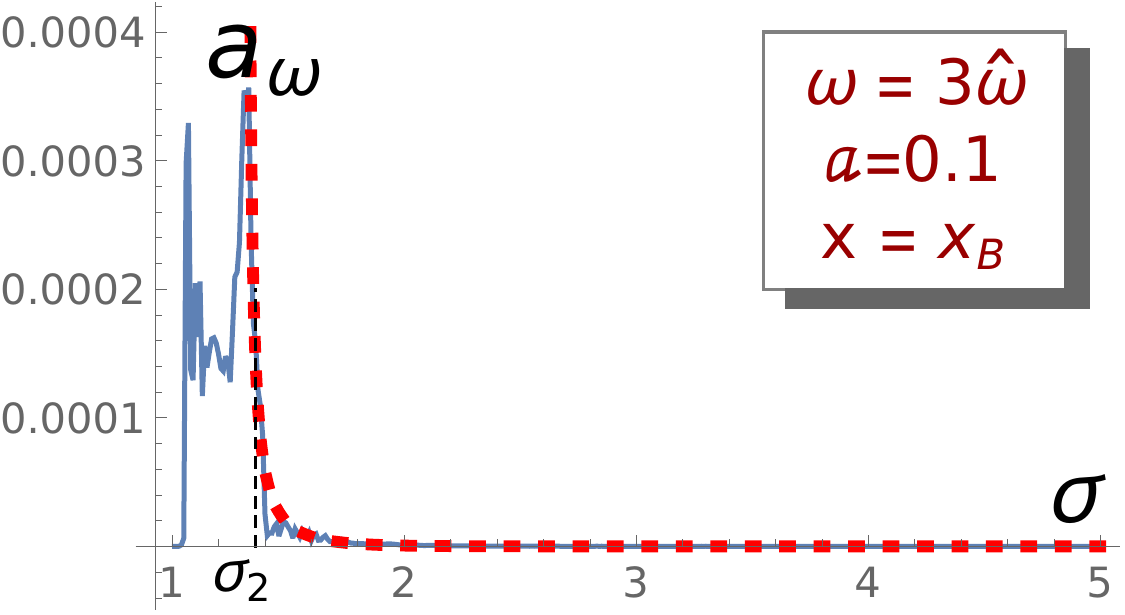}\qquad\includegraphics[width=0.31\textwidth]{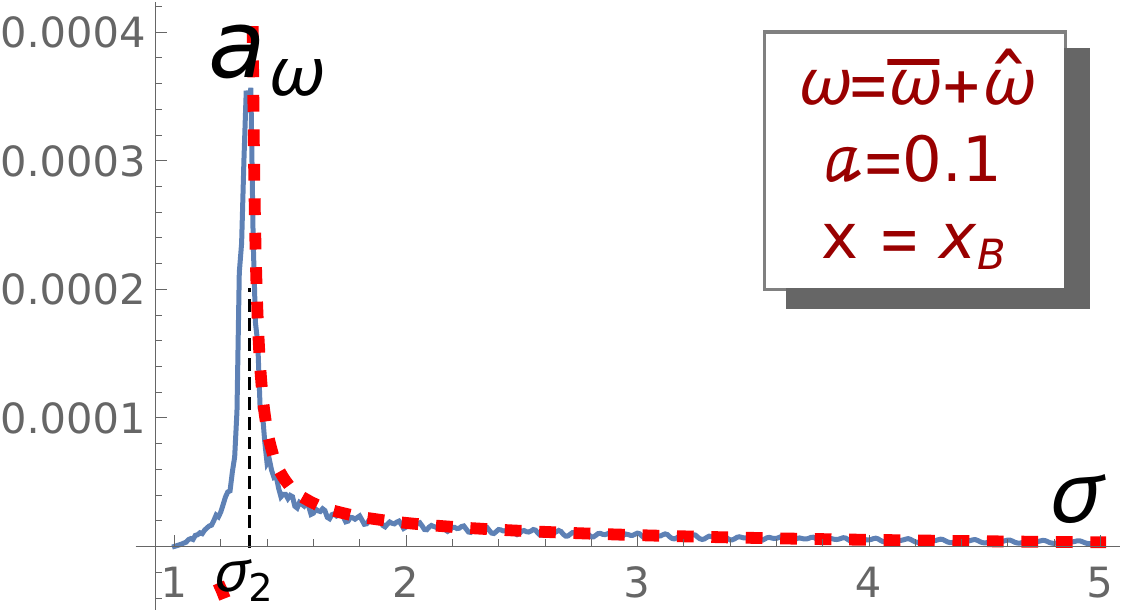}  }
	\caption{\small Comparison between the numerical and analytical results for the dependence on the parameter $\sigma$ of the amplitudes of the radiation emitted with frequencies $2\widehat{\omega}$, $3 \widehat{\omega}$ and $\overline{\omega}+\widehat{\omega}$. The red dashed curves represent the theoretical responses (\ref{radilong7})  and (\ref{radilong9}).} \label{fig:Manton01}
\end{figure}

The asymptotic behavior of orthogonal radiation can be obtained in a similar way for each frequency 
$\omega_\ell$ in \eqref{frequenciesell}. In this case, the second component of the orthogonal continuous eigenmodes is 
\[
\widehat{\eta}_{q_\ell} (x) = e^{iq_\ell x} \left( q_\ell+i \,\phi_K(x) \right)  ,
\]
with $q_\ell=\sqrt{\omega_\ell^2-\sigma^2}$  such that $|\omega_\ell| >\sigma$. The Wronskian associated to this case is given by 
\[
\widehat{W}_{q_\ell}= W[ \widehat{\eta}_{q_\ell}(x),  \widehat{\eta}_{-q_\ell}(x)] = 2 i q_\ell (q_\ell^2+1)  \, .
\]
Now the radiation emitted from the origin with frequency $\omega_\ell$ is determined by the general formula
\[
 \widehat{\eta}_{\omega_\ell}(x) = - \frac{ \widehat{\eta}_{-q_\ell}(x)}{\widehat{W}_{q_\ell}}  \int_{-\infty}^x  \widehat{\eta}_{q_\ell} (\xi) \, g_{\ell}(\xi)\ d\xi 
 - \frac{ \widehat{\eta}_{q_\ell}(x)}{\widehat{W}_{q_\ell}}  \int_x^\infty  \widehat{\eta}_{-q_\ell} (\xi) \, g_{\ell}(\xi) \ d\xi  \, ,
\]
which allows us to obtain the following asymptotic behavior for $ \widehat{\eta}_{\omega_\ell}(x)$:
\begin{eqnarray}
	 \widehat{\eta}_{3 \widehat{\omega}}(x) \!\!&\!\!\stackrel{x\rightarrow \infty}{\longrightarrow} \!\!&\!\! - e^{-iq_2x}\  \frac{3 \pi^2\,  \widehat{a}_0^3}{128(4\sigma^2 - 7)}  \,  \frac{q_2^3\, (q_2-i)}{i(q_2^2+1) \sinh\frac{\pi q_2}{2}}  \, , \hspace{0.7cm} q_2= \sqrt{8\sigma^2-9} \, , \\  [1ex]
	 \widehat{\eta}_{\overline{\omega}+\widehat{\omega}}(x) \!\!&\!\!  \stackrel{x\rightarrow \infty}{\longrightarrow} \!\!&\!\!  e^{-iq_3x}\ \frac{\pi^2 \,  \widehat{a}_0^3\,  (\sigma^2-1)}{32(4\sigma^2 - 7)}\,  \frac{q_3^3\, (q_3-i)}{i(q_3^2+1) \sinh\frac{\pi q_3}{2}}   \, , \hspace{0.5cm} q_3= \sqrt{2+2\sqrt{3}\sqrt{\sigma^2-1}}  \, . \label{xxx}
\end{eqnarray}
Note that the frequency $\omega_4=\overline{\omega}-\widehat{\omega}$ has not been considered in this calculation because in the regime $\sigma > \sigma_1$ its value is under the orthogonal continuous spectrum. Again, taking the imaginary part of the previous expressions, we obtain the physically relevant solutions, which are
\begin{eqnarray*}
     \widehat{\eta}_{3 \widehat{\omega}}(x,t) \!\!&\!\! \stackrel{x\rightarrow \infty}{\longrightarrow} \!\!&\!\! \frac{3 \pi^2 \,  \widehat{a}_0^3}{128(4\sigma^2 - 7)} \,\frac{q_2^3}{\sqrt{q_2^2+1} \sinh \frac{\pi q_2}{2}} \,   \sin (\omega_2 t -q_2 x+\delta_2)  ,
    \qquad
    \delta_2=\arctan \left( q_2\right)    ,\\   [1ex]
     \widehat{\eta}_{\overline{\omega}+\widehat{\omega}}(x,t) \!\!&\!\! \stackrel{x\rightarrow \infty}{\longrightarrow} \!\!&\!\!  -\frac{\pi^2 \,  \widehat{a}_0^3\,  (\sigma^2-1)}{32(4\sigma^2 - 7)} \frac{q_3^3}{\sqrt{q_3^2+1} \sinh \frac{\pi q_3}{2}} \, \,  \sin (\omega_3 t -q_3 x+\delta_3)  ,
    \qquad
    \delta_3= \arctan \left(q_3 \right),
\end{eqnarray*}
so the amplitudes of the radiation with frequencies $\overline{\omega}+\widehat{\omega}$ and $3 \widehat{\omega}$ are written as
\begin{eqnarray}
	 a_{3 \widehat{\omega}} = \frac{3 \pi^2 \,  \widehat{a}_0^3}{128(4\sigma^2 - 7)} \,\frac{q_2^3}{\sqrt{q_2^2+1} \sinh \frac{\pi q_2}{2}}  \, , \qquad 
 	a_{\overline{\omega}+\widehat{\omega}} =
 \frac{\pi^2 \,  \widehat{a}_0^3\, (\sigma^2-1)}{32(4\sigma^2 - 7)} \frac{q_3^3}{\sqrt{q_3^2+1} \sinh \frac{\pi q_3}{2}}  \, . \label{radilong9}
\end{eqnarray}
Figure \ref{fig:Manton01} shows the analytical results found at (\ref{radilong7}) and (\ref{radilong9}) for the amplitudes of the radiation emitted with frequencies $2\widehat{\omega}$, $3 \widehat{\omega}$ and $\overline{\omega}+\widehat{\omega}$, and  are compared with the numerical results shown in Section~\ref{Interaction}. It can be seen that the dependence of these amplitudes on the model parameter $\sigma$ is very well adjusted for the last two frequencies in the interval $\sigma > \sigma_1$. 
Furthermore, the divergences in the radiation amplitudes occur for the same values of the parameter $\sigma$.

\subsection{Amplitude decay law}

It is well known that the average power radiated in a period by a wave of the form $\eta=
A \cos(\omega t - q x + \delta)$ is $ \langle P\rangle =- \frac{1}{2}A^2 \omega q$, but as this radiation is emitted both to the right and to the left of the real line, then the power emitted would be doubled, that is, $ \langle P_{total} \rangle= -A^2 \omega q$. 
In the previous section we have seen that at distances far from the origin of coordinates in the first field radiation is emitted with frequency $2\widehat{\omega}$ while in the second field the frequencies are $3\widehat{ \omega} $ and $\overline{\omega}+\widehat{\omega}$.
Hence, rewriting the amplitudes described by the formulas \eqref{radilong7} and \eqref{radilong9} as
\begin{equation}
 a_{2 \widehat{\omega}}=\widehat{a}_0^2 A_{2 \widehat{\omega}} , \quad a_{3 \widehat{\omega}}= \widehat{a}_0^3 A_{3 \widehat{\omega}}, \quad a_{\overline{\omega}+ \widehat{\omega}}=\widehat{a}_0^3 A_{\overline{\omega}+ \widehat{\omega}} ,
\end{equation}
and taking into account that the radiation is emitted both from the right and from the left, we obtain that the average power radiated by the system is
\begin{equation} \label{potencia}
\langle P \rangle=\frac{d E}{d t}= -\left(\widehat{a}_0^4\, A_{2 \widehat{\omega}}^2 \,(2\widehat{\omega})  \, q_1  + \widehat{a}_0^6 \, A_{3 \widehat{\omega}}^2\,  (3 \widehat{\omega}) \, q_2   + \widehat{a}_0^6 \, A_{\overline{\omega}+ \widehat{\omega}}^2\,   (\overline{\omega}+ \widehat{\omega}) \, q_3\right).
\end{equation}

On the other hand, in the previous Section we have considered that the differential equation that describes the dynamics of $\widehat{a}$ is approximately
\begin{equation}
\widehat{a}_{tt} + \widehat{\omega}^2  \, \widehat{a} \approx 0,
\end{equation}
that is, we consider that the orthogonal discrete vibrational mode behaves as a harmonic oscillator at each point in space. Therefore, we are going to assume that the energy density of this vibration is given by the expression
\begin{equation}
 \mathcal{E}=\frac{1}{2} \widehat{\omega}^2 (\widehat{a}_0 \widehat{\eta}_D)^2.
 \end{equation}
Integrating this last expression over the whole space, we obtain that the corresponding total energy will be $E=\widehat{\omega}^2 \widehat{a}_0^2$. As for the discrete mode of the first field, it will not be considered because we are going to assume that we are working for very long times, in which this vibration has already been excited and, therefore, the energy of the orthogonal discrete mode will be inverted only in terms of exciting radiation. If we now take into account the formula \eqref{potencia} and neglect the sixth order terms, we arrive at the differential equation
 \begin{equation}
-\widehat{\omega}^2 \frac{d \widehat{a}_0^2}{d t} \approx \widehat{a}_0^4 A_{2 \widehat{\omega}}^2 2\widehat{\omega}  q_1,
\end{equation}
whose solution is 
 \begin{equation}\label{la432}
\widehat{a}_0^2(t) \approx \frac{\widehat{a}_0^2(0)}{\displaystyle 1+\frac{ 2q_1 A_{2 \widehat{\omega}}^2 }{\widehat{\omega}}\, \widehat{a}_0^2(0)\, t}.
\end{equation}
Function \eqref{la432} directly gives us the decay law of the orthogonal discrete mode amplitude which, as we saw in Figure~\ref{fig:simulacion}, shows a clear decrease in the value of this amplitude as time progresses.

In the next Section we will calculate another decay law for this same amplitude which, despite not being exactly the same, will have a very similar form to the one calculated here. 
This discrepancy in the amplitude decay of the discrete vibration modes also occurred in the study of the $\phi^4$ model in  \cite{Manton1997,Barashenkov2009,Barashenkov2018} and in fact the law obtained in this section has a very similar form to those calculated in these references.
Finally, we can highlight that, although we have not considered the sixth-order terms in \eqref{potencia}, they show that the orthogonal channel radiation influences the time evolution of $\widehat{a}_0$, although to a lesser extent than the radiation emitted in the parallel channel.

\section{Perturbative approach: Barashenkov and Oxtoby procedure}\label{BarashenkovOxtoby}

Barashenkov and Oxtoby employed a more formal approach using singular perturbation theories in \cite{Barashenkov2009} to address the same problem.
In this Section, the singular perturbation  theory will be applied to our problem following precisely the procedure described in \cite{Barashenkov2009}. As usual, the starting point is given by the equations (\ref{pde1}) and (\ref{pde2}). Taking into account the scenario described in Section~\ref{Interaction}, the following expansions of the fields about the kink are proposed
\begin{eqnarray}
\phi \!\!&\!\!=\!\!&\!\!
 \phi_0 \hspace{0.9cm} + \epsilon^2 \phi_2 \hspace{0.9cm}  + \epsilon^4 \phi_4 + \dots  \, , \nonumber\\   [1ex]
\psi \!\!&\!\!=\!\!&\!\!
 \hspace{0.7cm} \epsilon \psi_1 \hspace{0.9cm} + \epsilon^3 \psi_3 \hspace{1.2cm} + \dots  \, , \label{expansion}
\end{eqnarray}
where $\epsilon$ is a formal small real parameter. In addition to this expansion, a sequence of {\it stretched spaces and times\/} will also be used:
\begin{equation}
X_n \equiv \epsilon^n \, x  \, , \qquad T_n \equiv \epsilon^n \, t  . \label{slowvariables}
\end{equation}
In the limit $\epsilon\rightarrow 0$ the variables $X_n$ and $T_n$ become independent. The derivatives with respect to the original variables can also be expanded into the new ones as follows:
\begin{eqnarray}
	\frac{\partial}{\partial x}  \!\!&\!\!=\!\!&\!\!
 \partial_{X_0} + \epsilon \, \partial_{X_1} + \epsilon^2 \, \partial_{X_2} + \dots  \, , \hspace{0.8cm} \mbox{with} \hspace{0.5cm} \partial_{X_n} \equiv  \frac{\partial}{\partial X_n} \, ,  \label{dnx} \\    [1ex]
	\frac{\partial}{\partial t} \!\!&\!\!=\!\!&\!\!
 \partial_{T_0} + \epsilon \, \partial_{T_1} + \epsilon^2 \, \partial_{T_2} + \dots   \, , \hspace{1cm} \mbox{with} \hspace{0.5cm} \partial_{T_n} \equiv  \frac{\partial}{\partial T_n}  \, . \label{Dnt}
\end{eqnarray}
If we substitute the expressions (\ref{expansion})--(\ref{Dnt}) in (\ref{pde1})--(\ref{pde2}) we obtain a hierarchy of equations for the fields $\phi_k$ and $\psi_k$ for different orders in the $\epsilon$-expansion, which we will consider in some detail below.

\subsection{Zero order approximation: $\phi_0$}
In zero order approximation, the following relation
\[
\partial_{T_0,T_0} \phi_0 - \partial_{X_0,X_0} \phi_0 - 2\,\phi_0\, (1-\phi_0)^2 =0
\]
must be verified. As expected, the kink/antikink solution \eqref{kinkdef}
\begin{equation}
\phi_0 (X_0) = \pm\phi_K(X_0) = \pm \tanh X_0 \label{expansion0}
\end{equation}
meets the above condition and is considered the solution in this order.

\subsection{First order approximation: $\psi_1$}
In the first order approximation, the equation (\ref{pde1}) gives us the relation
\[
\partial_{T_0,T_1} \phi_0-\partial_{X_0,X_1} \phi_0  =0  \, ,
\]
which is checked automatically because as we have seen $\phi_0$ in (\ref{expansion0}) is independent of $T_0$, $X_1$ and $T_1$. On the other hand, the second Klein-Gordon equation (\ref{pde2}) leads to 
\[
\left(\partial_{T_0,T_0} - \partial_{X_0,X_0}  -2+\sigma^2 + 2\,\phi_0^2\right) \psi_1=0  \, ,
\]
which corresponds to the spectral problem for the orthogonal channel, see (\ref{operator}). Therefore,
\[
\psi_1 = \widehat{A}(X_1,\dots;T_1,\dots) \ e^{i\widehat{\omega} T_0} \ \widehat{\eta}_D (X_0)\, +\, {\rm c.c.} +  \widehat{\eta}(X_0,T_0)  \, ,
\]
where we recall that $ \widehat{\eta}_D(X_0) = {\rm sech}\, X_0$ and $\widehat{\omega} = \sqrt{\sigma^2 -1}$. As usual, the abbreviation ${\rm c.c.}$ indicates the complex conjugate of the previous expression, and
$$
   \widehat{\eta}(X_0,T_0) = \int_{-\infty}^\infty   \widehat{R} (q) \, e^{i\widehat{\omega}_q\, T_0}  \,\widehat{\eta}_q (X_0)\, dq + {\rm c.c.}
$$
is a wave packet traveling in the second field component, with orthogonal continuous eigenfunctions $\widehat{\eta}_q (x)$ defined on (\ref{continuous2}), and $\widehat{\omega}_q=\sqrt{\sigma^2 + q^2}$, as defined in \eqref{continuous2}. Taking into account that the problem we are dealing with only involves the initial excitation of the orthogonal shape mode, the field component $\psi_1$ is assumed to be 
\begin{equation}
\psi_1 = \widehat{A}(X_1,\dots;T_1,\dots) \, e^{i\widehat{\omega} T_0} \ {\rm sech}\, X_0 + {\rm c.c.} \label{expansion1}
\end{equation}

\subsection{Second order approximation: $\phi_2$}

In the second order approximation,  the following condition on $\psi_1$ 
\[
\partial_{T_0,T_1} \psi_1 - \partial_{X_0,X_1} \psi_1 =0
\]
is found. Plugging (\ref{expansion1}) into the previous relation we have
\[
\partial_{T_1} \widehat{A}  = 0  \, , \hspace{1cm}   \partial_{X_1} \widehat{A}  = 0  \, ,
\]
which implies that the amplitude $\widehat{A}$ does not depend on the slow variables $X_1$ and $T_1$, that is,
\[
\widehat{A} = \widehat{A} (X_2,\dots;T_2,\dots)  \, .
\]
On the other hand, the condition extracted in this order for the field component $\phi_2$ reduces to 
\begin{equation}
[\partial_{T_0,T_0} -\partial_{X_0,X_0}-2+6\phi_0^2] \, \phi_2 = -2 \, \phi_0\, \psi_1^2 = \overline{F}_2  \, ,\label{eqexpansion2}
\end{equation}
where 
\begin{eqnarray*}
\overline{F}_2 = - 4 |\widehat{A}|^2\, {\rm sech}^2 X_0 \tanh X_0 -2 \widehat{A}^2 \, {\rm sech}^2\, x_0 \tanh X_0 \, e^{i 2\widehat{\omega} T_0} + {\rm c.c.} 
= \overline{F}_2^{(0)} + \overline{F}_2^{(1)}e^{i 2\widehat{\omega} T_0} + {\rm c.c.}
\end{eqnarray*}
The solution of the homogeneous equation associated with (\ref{eqexpansion2}) can be written as
\begin{equation}
\phi_{2H}= C \, \overline{\eta}_0(X_0) + \overline{A} \, e^{i\, \overline{\omega} \, T_0} \, \overline{\eta}_D (X_0) + {\rm c.c} + \overline{\eta}(X_0,T_0)  \, , \label{expansion2}
\end{equation}
where $\overline{\eta}_0(X_0)={\rm sech}^2 X_0$, $\overline{\eta}_D(X_0) = {\rm sech}\, X_0 \tanh X_0$ and $\overline{\omega}=\sqrt{3}$ as explained in Section~\ref{MSTBmodel}. Besides, \begin{equation}
\overline{\eta}(X_0,T_0)=\int_{-\infty}^\infty   \overline{R}(q) e^{i \,\overline{\omega}_q\, T_0} \, \overline{\eta}_q(X_0)\ dq + {\rm c.c.}   \label{expansion2b}
\end{equation}
represents a wave packet evolving in the longitudinal channel, where $\overline{\eta}_q$ is given by formula \eqref{continuous1} with dispersion relation 
$\overline{\omega}_q = \sqrt{4+q^2}$. In the scenario described in Section~\ref{MSTBmodel}, the kink center is set to the origin, so the translational mode is not excited and the value $C=0$ can be assumed in (\ref{expansion2}). Based on the same arguments used for the orthogonal channel, the coefficients of the scattering eigenmodes are assumed to vanish: $\overline{R}(q)=0$ in (\ref{expansion2b}). Therefore, the equation (\ref{expansion2}) simplifies to
\begin{equation}
\phi_{2H}= \overline{A} \, e^{i\, \overline{\omega} \, T_0} \, {\rm sech}\, X_0 \tanh X_0 + {\rm c.c}   \, , \label{expansion222}
\end{equation}
A particular solution of the non-homogeneous linear differential equation (\ref{eqexpansion2}) can be found in the form 
\begin{equation}\label{phi_2}
\phi_{2P} = \phi_2^{(0)} + \phi_2^{(1)} e^{i\, 2 \widehat{\omega} \, T_0} +  {\rm c.c.} \, ,
\end{equation}
which leads to equations similar to (\ref{eqexpansion2}) for the variables $\phi_2^{(i)}$ associated to terms $\overline{F}_2^{(i)}$. The first of these equations reads 
\begin{equation}
[-\partial_{X_0,X_0} - 2 + 6\, \phi_0^2]\, \phi_2^{(0)} = -4 \, |\widehat{A}|^2 \, {\rm sech}^2 X_0 \tanh X_0  \, ,\label{eqexpansion2b}
\end{equation}
where it has been considered that $\phi_2^{(0)}$ does not depend on the slow time $T_0$. According to  Fredholm's alternative, inhomogeneous equations admit bounded solutions if and only if the functions 
$\overline{F}_2^{(i)}$ are orthogonal to the corresponding homogeneous solutions. In the case at hand, the homogeneous solution is $\phi_{2H}^{(0)}={\rm sech}^2 X_0$, which is orthogonal to $\overline{F}_2^{(0)}$. Therefore, the solution of (\ref{eqexpansion2b}) can be written as
\begin{equation}\label{phi_2^0}
\phi_2^{(0)} (X_0)= - |\widehat{A}|^2 X_0 \,{\rm sech}^2 X_0  \, .
\end{equation}
Note the factor $X_0 \,{\rm sech}\,X_0$ in the above expression, which does not oversize the previous term $\phi_0$ in the $\epsilon$-expansion, although they become larger than the difference $\phi_0\mp 1$ as $X_0\rightarrow \pm \infty$. This behavior could lead to non-uniformity of these expansions, see \cite{Barashenkov2009}. This problem can be solved by noting that $X_0\,{\rm sech}^2\,X_0$ can be expressed as the derivative of $\tanh(k X_0)$ with respect to $k$. This implies that
\[
\tanh [(1-|\widehat{A}|^2 ) X_0] \approx \tanh X_0 - |\widehat{A}|^2 X_0 \, {\rm sech}^2\, X_0 + o(|\widehat{A}|^2)  \, ,
\]
which means that $\phi_2^{(0)}$ is responsible for a variation of the kink size. 

On the other hand, the differential equation for $\phi_2^{(1)}$ is
\begin{equation}
 [-\partial_{X_0,X_0} -2 + 6 \, \phi_0^2 - (2\widehat{\omega})^2] \, \phi_2^{(1)} = -2 \, \widehat{A}^2 \, {\rm sech}^2 X_0 \tanh X_0  \, , \label{eqexpansion3}
\end{equation}
which must be solved under the symmetry conditions imposed by our problem, which in turn imply that the solution of (\ref{eqexpansion3}) must be an odd function in the variable $X_0$. This means that
\begin{equation}\label{phi_2^1}
\phi_2^{(1)} (X_0,X_2,\dots;T_2,\dots) = \widehat{A}^2  \,h(X_0;q_1)  + {\rm c.c.} \, , \hspace{0.6cm} q_1= 2\sqrt{\sigma^2-2}  \, ,
\end{equation}
where
\begin{eqnarray}
h(X_0;q) \!\!&\!\!=\!\!&\!\!
 - \frac{q^2-2}{4 (q^2+1)} \tanh X_0 +  e^{iqX_0} \frac{\pi q}{8(q^2+1)} \, {\rm cosech} \, \frac{\pi q}{2} \left( q^2-2+3iq \tanh X_0+3 \,{\rm sech}^2\,X_0 \right) \nonumber \\ [1ex]
\!\!&\!\! \!\!&\!\!
 + h_1(X_0,q) + h_1(X_0,-q)  \, ,
 \label{tururu}
\end{eqnarray}
with
\begin{eqnarray}
h_1(X_0,q) \!\!&\!\!=\!\!&\!\!
 \frac{-1}{8 (q^2+1)}\left(q^2-2 -3iq \tanh X_0 + 3\,{\rm sech}^2\,X_0 \right)  \nonumber\\ [1ex]
\!\!&\!\! \!\!&\!\!
 \hspace{1cm} \times \Big(  \, {}_2F_1{\textstyle[1,\frac{iq}{2},1+\frac{iq}{2},-e^{2X_0}]} 
 - \frac{q }{q-2 i} \, e^{2X_0}\, {}_2F_1{\textstyle[1,1+\frac{iq}{2},2+\frac{iq}{2},-e^{2X_0}]} \Big)
 \, .
\end{eqnarray}
Therefore, the final expression for the first component of the scalar field in the second order approximation $\phi_2$, taking into account \eqref{expansion222}, \eqref{phi_2^0} and \eqref{phi_2^1}, is given by 
\begin{equation}
    \phi_{2}= \overline{A} \, e^{i\, \overline{\omega} \, T_0} \, {\rm sech}\, X_0 \tanh X_0 \,   -\,  |\widehat{A}|^2 X_0 \,{\rm sech}^2 X_0 \,+\, \widehat{A}^2 \, e^{i\, 2 \widehat{\omega} \, T_0} \,h(X_0;q_1)\,+\, {\rm c.c.}
\label{expansion2c}
\end{equation}
The asymptotic behavior of this field is
\begin{equation}
\lim_{X_0\rightarrow \infty} \phi_2  \approx \frac{\pi \, \widehat{A}^2}{2} \,\, \frac{\sqrt{\sigma^2-2}}{\sinh (\pi \sqrt{\sigma^2-2})} \,\, \frac{\sqrt{\sigma^2-1}}{\sqrt{4\sigma^2-7}} \,\, e^{i\,(2\widehat{\omega} \, T_0 -q_1 X_0 + \delta_1) }  \, .
\label{labelitaita}
\end{equation}
From \eqref{labelitaita} it is immediate to see that, as in Section~\ref{MantonMerabet}, we find that the system emits radiation in the first component of the field with frequency $2\widehat{\omega}$. On the other hand, it can be verified that the radiation amplitude obtained in this section has a very similar shape to that calculated in the previous section. In fact, the only appreciable difference between the amplitudes obtained with both perturbative methods is that the one given in \eqref{labelitaita} is just twice the one that can be seen in \eqref{radilong7}.

\subsection{Third order approximation: $\psi_3$}

At third order we find the following condition for the field $\phi_2$
\[
-\partial_{X_0,X_1} \phi_2 + \partial_{T_0,T_1} \phi_2 =0  \, .
\]
If we substitute the solution (\ref{expansion2c}) into the previous equation, the amplitude of the longitudinal shape mode must comply with
\[
\partial_{X_1} \overline{A}(X_1,T_1)=\partial_{T_1} \overline{A}(X_1,T_1)=0  \, ,
\]
which means that
\[
\overline{A} = \overline{A} (X_2,\dots; T_2, \dots)  \, .
\]
On the other hand, the condition for the second field component at this order reads
\begin{equation}
\partial_{T_0,T_0} \psi_3 + [-\partial_{X_0,X_0} -2+\sigma^2 + 2\phi_0^2] \psi_3 =\widehat{F}_3  \, ,
\label{expansion3}
\end{equation}
where
\begin{eqnarray}
\widehat{F}_3 \!\!&\!\!=\!\!&\!\!
 e^{i\widehat{\omega} T_0} [-2\widehat{A}\, |\widehat{A}|^2 \, {\rm sech}\, X_0 (3 \, {\rm sech}^2 X_0 + 2 h(X_0) \tanh X_0) \nonumber  \\   [1ex]
\!\!&\!\! \!\!&\!\!
 - 2 i \, {\rm sech} \, X_0 (\widehat{\omega} \partial_{T_2} \widehat{A} - i \tanh X_0 \partial_{X_2}\widehat{A}) + 4|\widehat{A}|^2\widehat{A} \, X_0 \, {\rm sech}^3 X_0 \, \tanh X_0] + {\rm c.c.}  \nonumber  \\ [1ex]
\!\!&\!\! \!\!&\!\!
 +e^{3i\widehat{\omega} T_0} [-2 \widehat{A}^3 \, {\rm sech}\,X_0 ({\rm sech}^2 X_0 + 2 h(X_0) \tanh X_0)] + {\rm c.c.}   \nonumber  \\ [1ex]
\!\!&\!\! \!\!&\!\!
  + e^{i(\overline{\omega}+\widehat{\omega}) T_0} [-4 \overline{A} \, \widehat{A} \, {\rm sech}^2 X_0 \tanh^2 X_0] + {\rm c.c.}   \nonumber  \\ [1ex]
\!\!&\!\! \!\!&\!\!
  + e^{i(\overline{\omega}-\widehat{\omega}) T_0} [-4 \overline{A} \, \widehat{A}^* \, {\rm sech}^2 X_0 \tanh^2 X_0 ] + {\rm c.c.}   \nonumber  \\ [1ex]
\!\!&\!\!=\!\!&\!\!
 \widehat{F}_3^{ (0)}\, e^{i\widehat{\omega} T_0} +\widehat{F}_3^{ (1)}\, e^{3i\widehat{\omega} T_0} + \widehat{F}_3^{ (2)}\, e^{i(\overline{\omega}+\widehat{\omega}) T_0} + \widehat{F}_3^{ (3)}\, e^{i(\overline{\omega}-\widehat{\omega}) T_0}+ {\rm c.c.}
 \label{ddddd}
\end{eqnarray}
Exploiting the linearity of the equation (\ref{expansion3}) it can be assumed that 
\begin{equation}
\psi_3 =  {\psi}_3^{(0)} \, e^{i\widehat{\omega} T_0} +  {\psi}_3^{(1)} \, e^{3i\widehat{\omega} T_0} +   {\psi}_3^{  (2)} \,  e^{i(\overline{\omega}+\widehat{\omega}) T_0} +  {\psi}_3^{  (3)} \,  e^{i(\overline{\omega}-\widehat{\omega}) T_0} + {\rm c.c.} \label{expansion3b}
\end{equation}
If we substitute (\ref{expansion3b}) into the equation (\ref{expansion3}) we can determine the effect of every term $ \widehat{\psi}_3^{(i)}$ on the global solution. For example the ordinary differential equation which sets  $ \widehat{\psi}_3^{  (0)}$ has the form
\[ 
-\partial_{X_0,X_0}  {\psi}_3^{(0)} + (-2+\sigma^2+2\phi_0^2-\widehat{\omega}^2)  {\psi}_3^{(0)} = \widehat{F}_3^{(0)}  \, .
\]
From the Fredholm alternative to obtain bounded solutions the projection of $\widehat{F}_3^{(0)}$ on the homogeneous solution must be zero, that is, 
\begin{eqnarray*}
\int_{-\infty}^\infty dX_0 \, {\rm sech}\, X_0 \Big[-2 \widehat{A}\ |\widehat{A}|^2 \, {\rm sech}\, X_0 \, \left(3 \, {\rm sech}^2 X_0 + 2 \, h(X_0 \right) \tanh X_0) - \\ \hspace{1cm}- 2 \, i \, {\rm sech} \, X_0 \left(\widehat{\omega} \, \partial_{T_2} \widehat{A} - i \, \tanh X_0 \, \partial_{X_2} \widehat{A} \right)+ 4|\widehat{A}|^2\widehat{A} \, X_0 \, {\rm sech}^3 X_0 \, \tanh X_0 \Big] =0 \,  ,
\end{eqnarray*}
which leads to the condition
\begin{equation}
i \widehat{\omega} \, \partial_{T_2} \widehat{A} + \left( \frac{5}{3}+\xi(\sigma) \right)  \widehat{A} \ |\widehat{A}|^2 =0  \,  ,\label{ecuampliorto}
\end{equation}
where
\[
\xi (\sigma) =\int_{-\infty}^\infty dX_0 \ h(X_0; q) \, {\rm sech}^2 X_0 \tanh X_0  \, ,
\]
and $h(X_0)$ is given in \eqref{tururu}. 
Returning to the unscaled amplitude $\widehat{a}= \epsilon \widehat{A}$  of the orthogonal wobbling amplitude and the original time variable $t$, we can write (\ref{ecuampliorto}) as
\[
i\widehat{\omega}\ \frac{\partial \widehat{a}}{\partial t} + \left( \frac{5}{3}+\xi (\sigma) \right)  \widehat{a} \ |\widehat{a}|^2 = 0 \, ,
\]
which leads to the equation
\[
\frac{\partial |a|^2}{\partial t} = - \frac{2 \,{\rm Im}\,\xi(\sigma)}{\widehat{\omega}}\ |a|^4 , 
\]
such that
\[
|a(t)|^2 = \frac{|a(0)|^2}{1+ \frac{2 \,{\rm Im}\, \xi(\sigma)}{\widehat{\omega}} \ |a(0)|^2\, t} \, .
\]

The equation determining the field $\psi_3^{(1)}$ becomes
\begin{equation}
-\partial_{X_0,X_0} \psi_3^{(1)} + [-2+\sigma^2+ 2\phi_0^2- 9  \widehat{\omega}^2] \psi_3^{(1)} = \widehat{F}_3^{ (1)}   = -2 \widehat{A}^3 \, {\rm sech}\,X_0\ ({\rm sech}^2 X_0 +2 h(X_0) \tanh X_0)  \,  .\label{expansion3c}
\end{equation}
We recall that $\widehat{\eta}_q (x) = e^{iqx} (q+i \tanh x)$ and that the Wronskian associated to these solutions is given by $\widehat{W}_q = W[\widehat{\eta}_q (x), \widehat{\eta}_{-q} (x)]=2iq(q^2+1)$. With this notation the expression for $\psi_3^{(1)}$ extracted from (\ref{expansion3c}) can be written as
\[
\psi_3^{(1)}(X_0) = \frac{1}{\widehat{W}_{q_2}} \Big\{ \widehat{\eta}_{q_2} (X_0) \int \widehat{\eta}_{-q_2} (X_0)\, \widehat{F}_3^{ (1)}(X_0) \ dX_0 - \widehat{\eta}_{-q_2} (X_0) \int \widehat{\eta}_{q_2} (X_0) \widehat{F}_3^{ (1)} (X_0)\ dX_0 \Big\} \, ,  
\]
where $q_2=\sqrt{8\sigma^2-9}$ because $(3\widehat{\omega})^2 = 9 (\sigma^2 -1)$.

On the other hand, from \eqref{expansion3}--\eqref{ddddd} the equation for $\psi_3^{(2)}$ reads
\begin{equation}
-\partial_{X_0,X_0} \psi_3^{(2)} + [-2+\sigma^2+ 2\phi_0^2- (\overline{\omega}+\widehat{\omega})^2] \psi_3^{(2)} = \widehat{F}_3^{(2)} =  -4 \overline{A}\ \widehat{A} \, {\rm sech}^2 \,X_0 \tanh^2 X_0  \, .\label{expansion3d}
\end{equation}
In this case the solution is given by
\[
\psi_3^{(2)} (X_0,X_2,\dots;T_2,\dots) = \overline{A} \ \widehat{A}  \ g(X_0;q_3)   \, ,  
\]
where  $q_3$ is given in \eqref{xxx} and 
\[
g(X_0;q) = - \frac{q^2}{2 (q^2+1)} + \, {\rm sech}^2 X_0 +  \sum_{a=0,1} g_1\left((-1)^a X_0; q\right)  + \sum_{a,b=0,1} g_2 \left((-1)^a X_0; (-1)^b q \right)  \, ,  
\]
with
\begin{eqnarray*}
	g_1(X_0; q) \!\!&\!\!=\!\!&\!\!
 \frac{\pi\, q^3\, e^{iqX_0} }{8 (q^2+1) \sinh \, \frac{\pi q}{2}} \,  \Big( -i q + \tanh X_0  \Big)  \, ,  \\  [1ex]
	g_2(X_0; q) \!\!&\!\!=\!\!&\!\!
 \frac{q^2}{8 (q^2+1)}\left(-iq+ \tanh X_0 \right)  \\  [1ex]
	&& \hspace{1cm} \times \Big( \, {}_2F_1{\textstyle[1,\frac{iq}{2},1+\frac{iq}{2},-e^{-2X_0}]} - \frac{q }{q-2 i} \, e^{-2X_0}\, {}_2F_1{\textstyle[1,1+\frac{iq}{2},2+\frac{iq}{2},-e^{-2X_0}]} \Big)   \, .  
\end{eqnarray*}
We have that the asymptotic behaviour for $X_0\rightarrow \infty$ is
\begin{equation}
     \psi_3^{(2)}  \quad \!\! \!\!\stackrel{X_0\rightarrow \infty}{\longrightarrow} \!\! \!\! \quad  \overline{A} \ \widehat{A}\ \frac{\pi q_3^3}{4 (-1+i q_3) \sinh \frac{\pi q_3}{2}} e^{i\,((\overline{\omega}+\widehat{\omega}) \, T_0 -q_3 X_0 ) } + {\rm c.c.} \label{rad3omega}\, 
\end{equation}
As we can see, this term corresponds to a radiative term, as occurred in Section~\ref{MantonMerabet}. In the present case, the amplitude obtained is different, although the qualitative behavior obtained from \eqref{rad3omega} is the same as that obtained can extract from \eqref{radilong9}.

\subsection{Fourth order approximation: $\phi_4$}

At fourth order we find the following condition for the field $\phi_4$
    \begin{eqnarray}
&\hspace{-11cm}\partial_{T_0,T_0} \phi_4 + [- \partial_{X_0,X_0} -2+ 6 \phi_K^2 ] \phi_4 = -(6\phi_0 \phi_2^2 + 2 \phi_2 \psi_1^2 + 4 \phi_0\psi_1\psi_3-\\ 
\hspace{7cm}& -2\partial_{X_0,X_2}\phi_2-\partial_{X_1,X_1}\phi_2 +2 \partial_{T_0,T_2}\phi_2 + \partial_{T_1,T_1}\phi_2) = \overline{F}_4 .
    \nonumber
\end{eqnarray}

As in previous sections, the non-homogeneous term of this differential equation will present terms with different frequencies, so that
\begin{eqnarray}
   & \hspace{-4cm} \overline{F}_4=\overline{F}_4^{(0)}+    \overline{F}_4^{(1)} e^{ i  \overline{\omega} T_0}  +   \overline{F}_4^{(2)} e^{ 2 i  \widehat{\omega} T_0}                                                                            +  \overline{F}_4^{(3)} e^{ 2i  \overline{\omega} T_0} + \label{cuartoorden} \\
  &\hspace{4cm} + \overline{F}_4^{(4)}e^{4 i  \widehat{\omega} T_0} +\overline{F}_4^{(5)} e^{2 i  \widehat{\omega} T_0+ i \overline{\omega} T_0 } +\overline{F}_4^{(6)}e^{2 i  \widehat{\omega} T_0- i \overline{\omega} T_0 } + {\rm c.c.} \nonumber
\end{eqnarray}
We will now focus only on the  term with frequency $\overline{\omega}$, as it will be used to find the decaying law for $ \overline{A}$. For this frequency, the corresponding differential equation will be the following:
\begin{eqnarray*}
\!\!&\!\! \!\!&\!\! -\overline{\omega}^2 \phi_4^{(1)}+[- \partial_{X_0,X_0} -2+ 6 \phi_0^2 ] \phi_4^{(1)}=-4|\widehat{A}|^2 \overline{A}\, {\rm sech^3} \, X_0\, \tanh \, X_0 + 12 X_0\,  |\widehat{A}|^2 \,\overline{A} \, {\rm sech^3}\, X_0 \tanh^2 X_0  -\\   [1ex]
\!\!&\!\! \!\!&\!\! 
-4|\widehat{A}|^2 \overline{A}\, {\rm sech} \, X_0\, \tanh X_0 \, g(X_0) - 2 i\, \overline{\omega} \, {\rm sech} \, X_0\, \tanh X_0 \partial_{T_2} \overline{A} + 2 \, {\rm sech^3} \, X_0 \, \partial_{X_2} \overline{A}- \\   [1ex]
\!\!&\!\! \!\!&\!\! 
-2 \, {\rm sech} \, X_0\, \tanh^2 (X_0)\, \partial_{X_2} \overline{A}=  \overline{F}_4^{(1)}(X_0).
\end{eqnarray*}
From Fredholm's alternative we know that in order to obtain bounded solutions, in the previous equation the solution of the homogeneous part must be orthogonal to the inhomogeneous term. This results in the following condition
\begin{equation}
\int_{-\infty}^{\infty}d X_0\left({\rm sech} \, X_0 \tanh X_0 \overline{F}_4^{(1)}(X_0)\right)=0,
\end{equation}
which leads us to the differential equation
\begin{equation}\label{ecamplitudpar}
- 5 i \overline{\omega}\, \partial_{T_2} \overline{A} +(3-15 \mu(\sigma) ) \overline{A}|\widehat{A}|^2=0,
\end{equation}
where 
\begin{equation}
\mu(\sigma)=\int_{-\infty}^{\infty}d X_0\, g(X_0) {\rm sech^2} \, X_0 \tanh^2 X_0.
\end{equation}

If $\widehat{a}=\epsilon \widehat{A}$ and $\overline{a}=\epsilon\overline{A}$, then
\begin{equation}\label{ecdifampliort2}
\frac{\partial |\overline{a}|^2}{\partial t}= |\overline{a}|^2|\widehat{a}|^2 \left(\frac{-6 \, {\rm Im}\,\mu(\sigma)}{\overline{\omega}}\right).
\end{equation}
Finally, the solution of \eqref{ecdifampliort2} leads to the time evolution law for $\overline{a}$ being
\begin{equation}
|\overline{a}(t)|^2= |\overline{a}(0)|^2 (\overline{\omega}\widehat{\omega})^{\frac{3\, {\rm Im} \, \mu(\sigma)\widehat{\omega}}{{\rm Im} \, \xi(\sigma)\overline{\omega}}}\left( (2 |\widehat{a}(0)|^2{\rm Im}\, \xi(\sigma)\, t +\widehat{\omega})\overline{\omega}\right)^{-\frac{3\, {\rm Im} \, \mu(\sigma)\widehat{\omega}}{{\rm Im} \, \xi(\sigma)\overline{\omega}}}.
\end{equation}
This procedure has some similarities with the one carried out in the previous section, since, for example, the differential equations \eqref{ecuampliorto} and \eqref{ecamplitudpar} present a very similar form to each other.
However, there are also very important differences. For example, in this case we can see how $\overline{A}$ is affected by $\widehat{A}$, while in the calculation of $\widehat{A}$ something similar does not happen. This is reasonable since we are initially only exciting the discrete mode of the second field, so it makes sense that $\widehat{A}$ dictates the behavior of $\overline{A}$.
An advantage of this perturbative theory with respect to the Manton and Merabet method is that we have managed to reach fourth order. It is important to note that in  \eqref{cuartoorden} the frequency terms $4\widehat{\omega}$ and $2\widehat{\omega}+\overline{\omega}$ appear, which correspond to the extra radiative terms that we found in Figure~\ref{fig:espectro8a}  .

\section{Concluding remarks}\label{conclusion}

In this paper we have investigated the interaction between longitudinal and orthogonal shape modes in a two-component real scalar field theory. We have shown that orthogonal shape mode excitation in MSTB model trigger longitudinal excited modes. The energy transfer mechanism between shape modes and its impact on the fractal structure found in the  final versus initial velocity diagram is not completely addressed in the literature. The present investigation brings numerical and analytical novelty results from a different perspective that shed light onto the energy transfer between excited modes both in one-component and two-component real scalar field models.

A valuable piece of information for investigating the interaction between different component shape modes in the MSTB model is that the value $\sigma$ of dictates the excitation amplitude of longitudinal shape mode. Besides, results have shown to be almost independent of the initial amplitude excitation of the orthogonal shape mode. We describe the main numerical simulation outcomes where initial values of  $\sigma$ and the point where spectral distribution for time series is evaluated determine the observed frequencies and amplitudes of both field excited bound states. The spatial points where we measure the eld excitation is important to differentiate whether the result found at a point $x$ is a localized fluctuation or a shape mode vibration.

In order to achieve a global understanding of the excitation process of different shape modes the
spectral analysis using a fast Fourier transform algorithm was carried out for  $\sigma$ values in the range $(1,5]$ with steps of $\Delta \sigma= 0.01$ and with a  fix value $a = 0.1$ for the initial amplitude of the orthogonal shape
mode. From numerical simulation results we have observed that:

\begin{itemize}
\item
For any value of  $\sigma$ both vibrational and radiation mode of the wobbling kink are triggered by the initial excitation on orthogonal  field component shape mode. This demonstrate that the energy stored in orthogonal  field component can freely excite its other component excited modes.
\item
Frequencies $2\widehat{\omega}$ and $\overline{\omega}$ propagate at $x_M$ in the first field spectral analysis. At this point, the
dominant frequencies are $2\widehat{\omega}$ for $\sigma<\sigma_2$ and $\overline{\omega}$ for $\sigma>\sigma_2$. The maximal value of wobbling
excitation and minimal orthogonal shape mode amplitude occurs for $\sigma=\sigma_2$ where these frequencies
are resonant.
\item
For $\sigma>\sigma_1$ we found radiation waves in the first field component vibrating with $2\widehat{\omega}$ at $x_B$, contrary to the case of usual $\phi^4$ kink where radiation is emitted with frequency $2\overline{\omega}$. The wobbling kink radiation mode
$\overline{\omega}_c$  is also found although its amplitude is small. The maximum longitudinal
excitation occurs for $\sigma=\sigma_1$ where   $\overline{\omega}_c$ and   $2\widehat{\omega}$ are resonant.

\item
As the value of $\sigma$ increases, longitudinal radiation frequency at $x_B$ approaches to the value  $\overline{\omega}_c$. This can be partly explained by the fact that higher frequencies are more difficult to be excited. It's important to remark that the orthogonal vibrational mode threshold grows as  $\sigma$ increases.

\item
In the spectral analysis for orthogonal radiation mode at  $x_B$ we found frequencies 
$\widehat{\omega}_c$, $3 \widehat{\omega}$ and $\widehat{\omega}+\overline{\omega}$. This clearly indicates that not only excitation on orthogonal modes can trigger longitudinal
excited modes but the reverse situation also occurs.
\end{itemize}

In an analytical approach some of the previous mentioned novelty results have been explained
by using two different perturbation expansion techniques. In one hand, Manton and Merabet's
approach has well fitted numerical simulation outcomes for dependence of the orthogonal radiation
amplitude on the model parameter  $\sigma$. On the other hand, Barashenkov and Oxtoby's procedure have ...

The interaction between longitudinal and ortogonal excited modes shown in this paper present
some fascinating behaviors and novelty results that open new possibilities for further works. A
fascinating idea to unveil more features in the energy transfer mechanism between shape modes is
to construct a new two-component scalar  field model where both component presents more bound
states above the zero mode in order to investigate how energy flows from one to other superiors
excited modes. Moreover, the present research expand new prospects to investigate the scattering
between excited topological defects in a two-component real scalar field model.

\section*{Acknowledgments}

This research was supported by Spanish MCIN with funding from European Union NextGenerationEU (PRTRC17.I1) and Consejeria de Educacion from JCyL through QCAYLE project, as well as MCIN project PID2020-113406GB-I0.
A. Alonso-Izquierdo acknowledges  the Junta de Castilla y Le\'on for financial support under grant SA067G19.  This research has made use of the high performance computing resources of the Castilla y Le\'on Supercomputing Center (SCAYLE),
financed by the European Regional Development Fund (ERDF).

\end{document}